\documentclass[reprint, superscriptaddress]{revtex4-2}

\usepackage{graphicx}
\usepackage{url}
\usepackage{amsmath,enumitem}
\usepackage{mathtools}
\usepackage{amssymb}
\usepackage[dvipsnames]{xcolor}
\usepackage{seqsplit}
\usepackage{url}

\newcommand{\hash}[1]{{\ttfamily\seqsplit{#1}}}
\usepackage{xcolor}
\definecolor{amaranth}{rgb}{0.9, 0.17, 0.31}

\usepackage{xr}

\makeatletter
\newcommand*{\addFileDependency}[1]{
  \typeout{(#1)}
  \@addtofilelist{#1}
  \IfFileExists{#1}{}{\typeout{No file #1.}}
}
\makeatother

\begin{document}

\title{The world-wide waste web\\}

\author{Johann H. Mart\'inez}
\affiliation{Instituto de F\'{\i}sica Interdisciplinar y Sistemas Complejos IFISC (CSIC-UIB), 07122 Palma de Mallorca, Spain} \affiliation{Department of Biomedical Engineering, Universidad de los Andes, Cr 1 18A-12, Bogot\'a, Colombia.}

\author{Sergi Romero}\affiliation{Instituto de F\'{\i}sica Interdisciplinar y Sistemas Complejos IFISC (CSIC-UIB), 07122 Palma de Mallorca, Spain}

\author{Jos\'e J. Ramasco}\affiliation{Instituto de F\'{\i}sica Interdisciplinar y Sistemas Complejos IFISC (CSIC-UIB), 07122 Palma de Mallorca, Spain}

\author{Ernesto Estrada}
\email[]{Email: estrada@ifisc.uib-csic.es}
\affiliation{Instituto de F\'{\i}sica Interdisciplinar y Sistemas Complejos IFISC (CSIC-UIB), 07122 Palma de Mallorca, Spain}
\affiliation{Institute of Mathematics and Applications, University of Zaragoza,
Pedro Cerbuna 12, Zaragoza 50009, Spain; ARAID Foundation, Government
of Aragon, Spain.}

\begin{abstract}
\section*{Abstract}
Countries globally trade with tons of waste materials every year, some of which are highly hazardous. This trade admits a network representation of the world-wide waste web, with countries as vertices and flows as directed weighted edges. Here we investigate the main properties of this network by tracking 108 categories of wastes interchanged in the period 2001-2019. Although, most of the hazardous waste was traded between developed nations, a disproportionate asymmetry
existed in the flow from developed to developing countries.
Using a dynamical model, we simulate how waste stress propagates
through the network and affects the countries. We identify 28 countries with low Environmental Performance Index that are at high risk of waste congestion. Therefore, they are at
threat of improper handling and disposal of hazardous waste. We find evidence of pollution
by heavy metals, by volatile organic compounds and/or by
persistent organic pollutants, which are used as chemical fingerprints, due to the improper handling of waste in several of these countries.
\end{abstract}


\maketitle 

\section*{Introduction}

Globally, 7-10 billion tonnes of waste are produced annually \cite{wilson2015waste,chen2020world},
including 300-500 million tonnes of hazardous waste (HW)--explosive,
flammable, toxic, corrosive, and of biological risk \cite{akpan2020hazardous,TheWorldCount}.
About 10\% \cite{krueger2001basel} of this HW is traded through
the \textit{world-wide waste web} (W4).
The W4 is a network formed by the legal trading of waste, where countries are represented as nodes and the flow of materials are encoded as  weighted  directed  links (edges).  The  forces  that  impulse  international trade of HW are complex and multifactorial, and generally involve economic, geographic, socio-political and environmental  aspects  difficult  to  disentangle. As a consequence, the  volume  of  HW  traded through the W4 in the last 30 years has grown by 500\% \cite{kellenberg2015economics} and will continue to grow \cite{kaza2018waste}, creating serious legal \cite{walters2020waste}, economic \cite{kellenberg2015economics}, environmental \cite{balayannis2020toxic} and health \cite{fazzo2017hazardous} problems at  global  scale.

It is frequently claimed that the global trade of HW is mainly a waste
flow from developed to developing countries \cite{sonak2008shipping,krueger2001basel}.
Although other studies point out to a more complex picture with important
contributions of the South-South trade \cite{lepawsky2015we} as well as
of South-North exports \cite{moore2018undermining}.

From an economic perspective waste trade may offer benefits to both
types of countries \cite{balayannis2020toxic}. Developed countries
would benefit from cheaper disposal costs in developing nations and
avoiding increasing resistance to HW disposal facilities in their
territories. Developing countries would gain access to cheap raw materials
by recycling wastes, rocketing production and employment. This would
be a win-win situation if it were not because many of the importer
nations are highly indebted countries with very poor track records
of waste management and environmental performance \cite{walters2020waste}.
Additionally, as revealed by several high profile cases \cite{klenovvsek2011international},
the situation is aggravated by illegal HW trafficking to, and dumping
in, developing countries \cite{favarin2020global}.

To address the problems of HW, the United Nations created in 1989 the Basel Convention
(BaC) on the Control of Transboundary Movement of Hazardous Wastes
and their Disposal \cite{Basel}. The Convention has as mandate the monitoring of global waste trading. Countries self-report the amount of imported and exported waste, and its origin/destination. Some type of waste as radioactive materials are excluded from the reports. In its more than 30 years, BaC has
revealed the difficulties to obtain accurate information regarding
the magnitude and direction of global HW flows \cite{krueger2001basel,yang2020trade}.
The information recorded by the BaC on waste trade does not contain information on illegal trade. However, it constitutes
the most reliable information for building a map of the W4, which
is vital to understand how the flows of HW are organized at global
and local scales. This analysis is necessary for efficiently managing
the transboundary HW trade and implementing more effective measures
for its better management and control.

Here we rely on data reported by countries and territories on their trade of 108
categories of HW during the years 2001-2019, except for 2010, for which data is not available. This data is the most
complete information about transboundary waste trade at the BaC database.
By merging these categories into seven classes of waste, we study the
trade networks that account for the legal flow of HW in the world.
First, we analyze the global characteristics of these networks. By considering the relation
between the simulated risk of waste congestion and countries' environmental
performance, we analyze the potential risks of improper handling and
disposal of HW by individual nations. Finally, we identify ``chemical
fingerprints'' that reveal the impact of improper handling and disposal
of HW on the environment and human health on 28 countries identified
at high risk.

\begin{figure*}
\begin{centering}
\includegraphics[width=18cm]{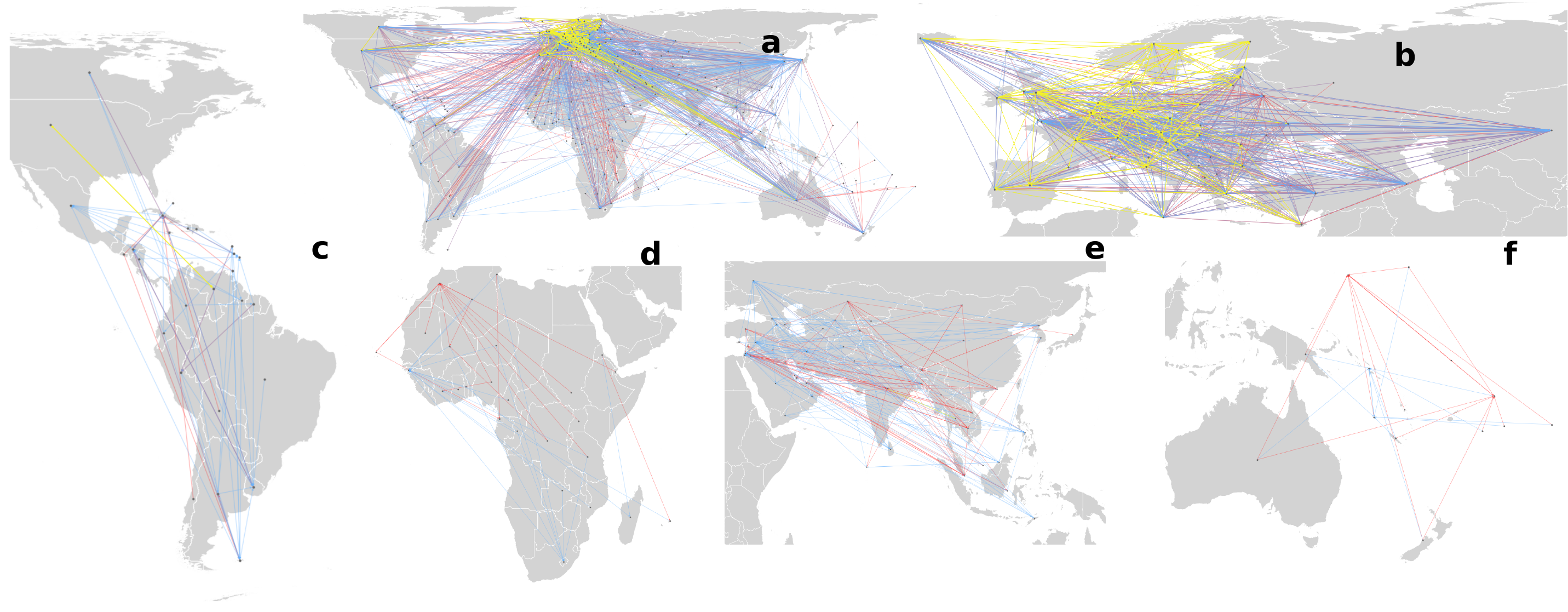}
\par\end{centering}
\caption{\textbf{The world-wide waste web}. Superposition of the W4 networks
of types I (red edges), type II (blue edges) and type III (yellow
edges) of waste, where the nodes represent the countries which traded
the corresponding waste in the years 2001-2019. The direction of the
edges indicates the flow from exporter to importer as reported at
the BaC database. A view of the global network in \textbf{a}, with zooms for the local networks of Europe in \textbf{b}, the Americas \textbf{c}, Africa \textbf{d}, Asia \textbf{e} and Oceania \textbf{f}. Map tiles by Bjorn Sandvik, under CC BY-SA 3.0 available at http://thematicmapping.org/downloads/world\_borders.php.} \label{wastes percentage}
\end{figure*}

\section*{Results}
\subsection*{Theoretical modeling approach}
We consider that the change of the amount of waste $w_{i}$ in a country
$i$ over time obeys a logistic model:
\begin{equation}
\dfrac{dw_{i}\left(t\right)}{dt}=\dot{w}_{i}\left(t\right)=\beta\left(1-w_{i}\left(t\right)\right)\sum_{j=1}^{n}A_{ij}w_{j}\left(t\right),\label{eq:logistic}
\end{equation}
where $\beta$ quantifies how quickly waste grows per amount of waste
already in the country and $A_{ij}$ is the normalized amount of waste
exported by country $i$ to country $j$. In a period of time, let
say from January to December of a given year, the amount of waste
accumulated at a given country growth exponentially at earlier times
if there is little waste at the beginning and enough processing resources
in that country (including its exports of waste). But when the amount
of waste gets large enough, processing resources start to get congested,
slowing down the growth rate. Eventually, it will level off, completing
a characteristic S-shaped curve. The amount of waste at which it levels
off, represents the maximum amount of waste that this particular country
can support, and it is called its carrying capacity. We say that a
country which has reached its carrying capacity is congested or saturated.
We consider here that a country $i$ in the year $t$ has a carrying
capacity equal to the total amount of waste traded by it, i.e., imported
and exported, that year. We normalize this carrying capacity for every
country such that it is equal to one. In this way all the countries
will congest when $\lim_{t\rightarrow\infty}w_{i}\left(t\right)=1$.
Let us call congestion time $t_{s}\left(i\right)$ to the time at
which $w_{i}\left(t_{s}\left(i\right)\right)=1$. Some countries will
reach this congestion time earlier than others. Suppose that a country
reaches its carrying capacity in a simulation time equivalent to February
in the real time. Then, because this situation is physically implausible
we consider that such country is at risk of getting over-congested
of waste during the rest of the year. In other words, we consider
the congestion time as a proxy of the risk at which a country is exposed
by the international trade of waste.

\begin{figure*}
\begin{centering}
\includegraphics[width=16cm]{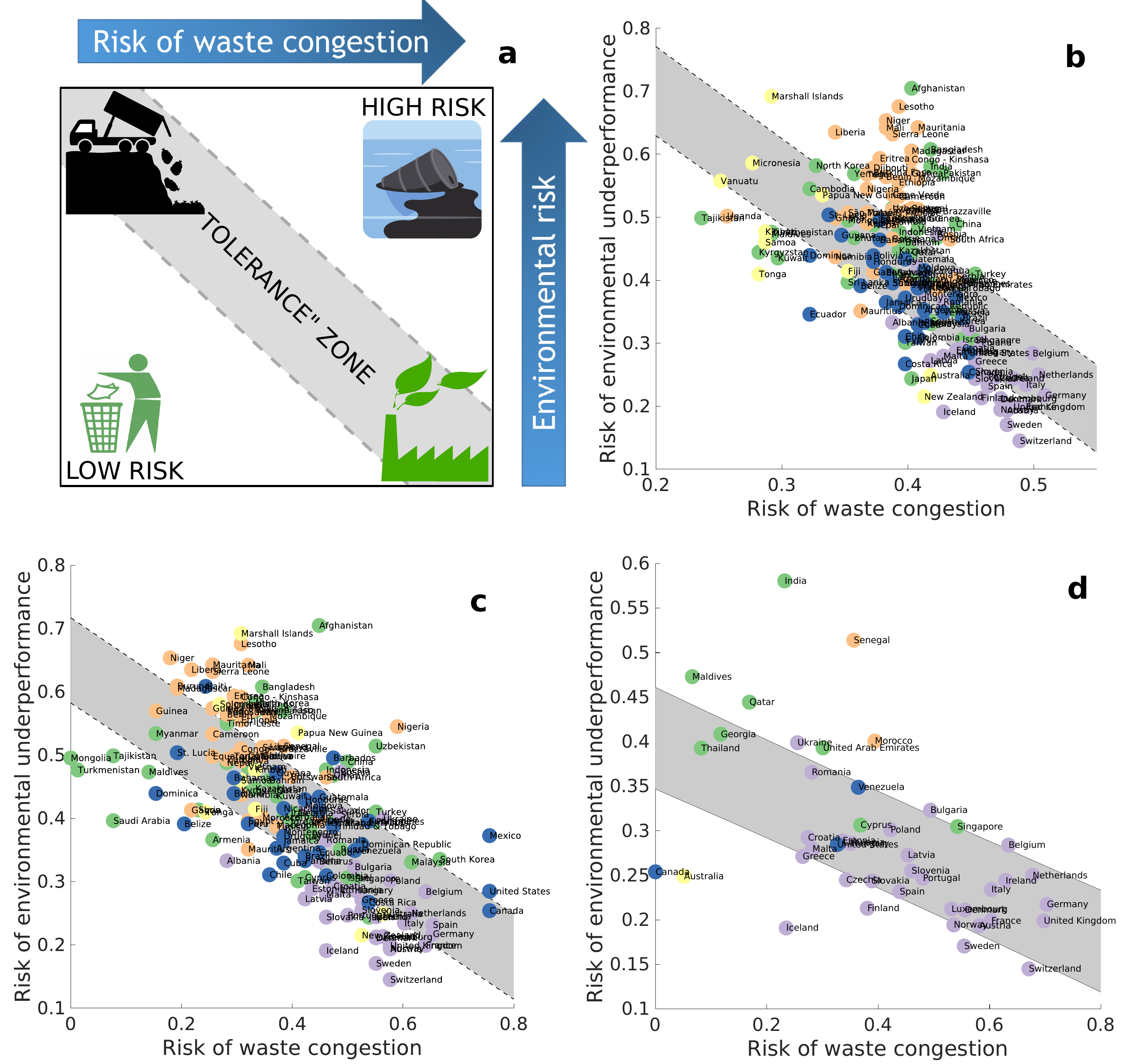}
\par\end{centering}
\caption{\textbf{Potential Environmental Impact of Waste Congestion} \textbf{(PEIWC)}.
Plot of the risk of waste congestion versus the environmental performance
\textbf{a} indicating the central region of ``tolerance'' where countries
process waste with relatively low environmental and human health impacts.
The tolerance zone is defined here by the upper and lower 50\% prediction
bounds for response values associated with the linear regression trend
between the two risk indices. Countries over the tolerance zone are
at high risk of improper handling and disposal of wastes (HRIHDW). \textbf{b}-\textbf{d} Illustration of
the PEIWC for wastes of types I-III, respectively. Nodes are colored by the continent to which the country belongs to: blue (Americas), purple (Europe), yellow (Africa), green (Asia). Icons of panel \textbf{a} were obtained from https://www.pdclipart.org/ under CC Public Domain.}
\label{Potential Environmental Impact}
\end{figure*}

The first time derivative $\dot{w}_{i}\left(t\right)$ in the logistic
model accounts only for properties of this change in an infinitely
small neighborhood of the considered time $t$. However, the change of
the amount of waste in a given country at $t$ depends on the
changes of input of wastes on finite (or infinite) time interval of
the past. This is known as \textit{nonlocality by time} or \textit{dynamic memory} \cite{TARASOV2018157,du2013measuring}.
That is, the saturation of waste by a given country in a year depends
not only on what it trades this year, but also on the waste ``accumulated''
by trading in the past by this country. In order to account for this
temporal nonlocality or memory of the process we will replace here
the first time-derivative by a fractional one. We use the Caputo time-fractional
derivative, which when applied to a given function $f\left(t\right)$
is defined as \cite{mainardi2010fractional}
\begin{equation}
D_{t}^{\alpha}f\left(t\right)\coloneqq\dfrac{1}{\varGamma\left(\kappa-\alpha\right)}\int_{0}^{t}\dfrac{f^{\left(\kappa\right)}\left(\tau\right)d\tau}{\left(t-\tau\right)^{\alpha+1-\kappa}},
\end{equation}
where $\kappa=\left\lceil \alpha\right\rceil $ is the ceiling function applied to $\alpha$. Here, we consider $0<\alpha\leq1$ and $\kappa=1$.

In order to understand how the Caputo fractional derivative captures
	the past history of the system \cite{caputo2021}, let us write it as:
\begin{equation}
^{C}D_t^{\alpha}\, =\dfrac{1}{\varGamma\left(1-\alpha\right)}\int_{0}^{t}\bigg[ \dfrac{1}{(t-s)^\alpha}\bigg]\, \dfrac{d f\left(s\right)}{ds}.
\end{equation}

The term in the squared parenthesis inside the integral represents a weight given to the standard derivative. Then, we can consider the initial time of the integration as the remote past, while $t$ is the present. Therefore, when we consider a time $k$ far apart from the present $t$, the term $t-s$ is relatively large, thus if $\alpha =1$ the term $(t-s)^{-\alpha}$ vanishes, which indicates no contribution from this remote past time. However, if $\alpha \ll 1$ such term is different from zero and the derivative $\dfrac{df(s)}{ds}$ receives a weight different from zero for the contribution of this remote past time. In closing, the smaller the value of $\alpha$ the larger the contribution from the remote past \cite{caputo2021}. That is, the system remembers its past history. If $\alpha=1$, only the present time is considered by the derivative, which indicates no memory from the past history of the system.

\begin{figure*}
\begin{centering}
\includegraphics[width=18cm]{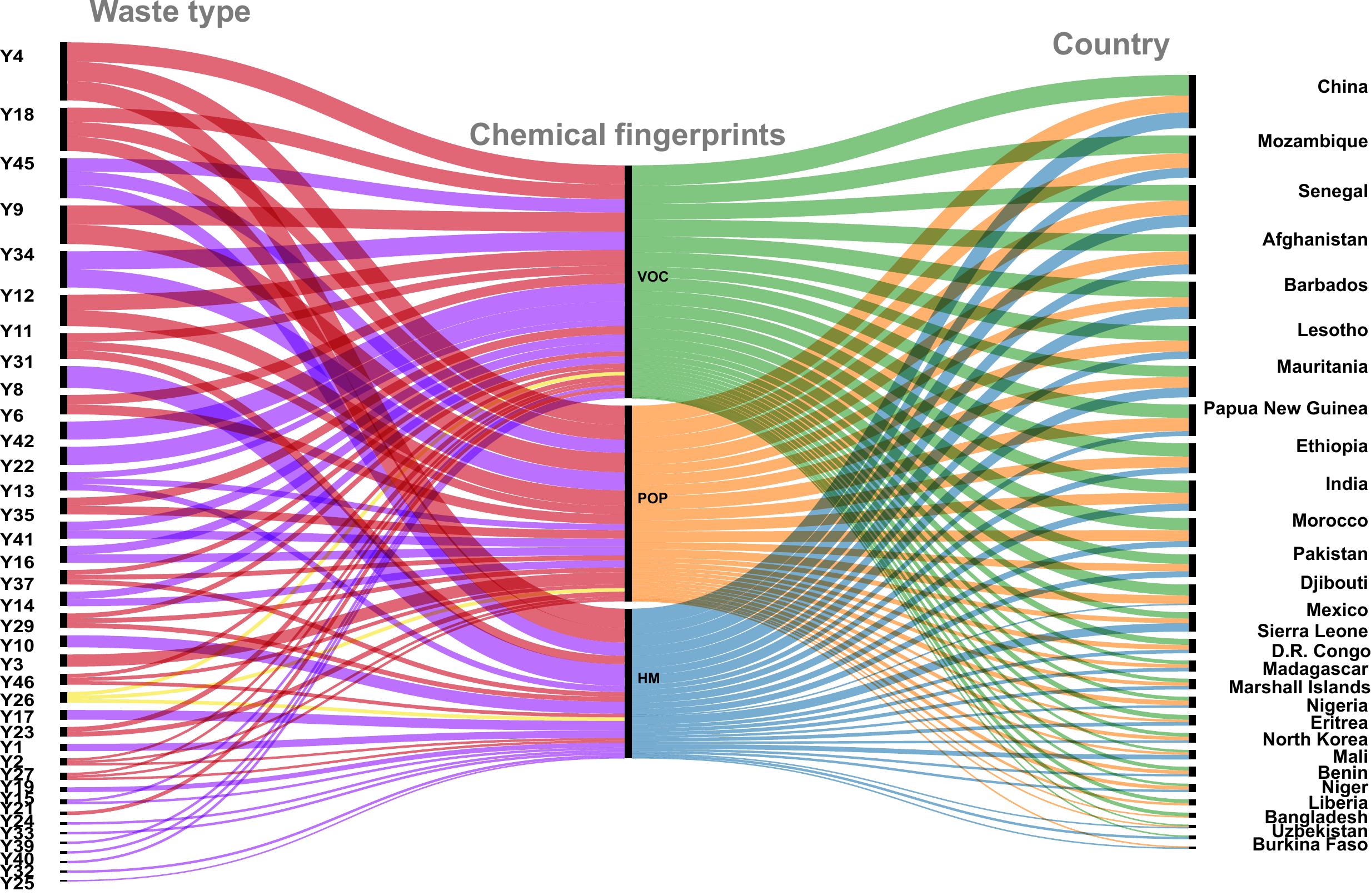}
\par\end{centering}
\caption{\textbf{Chemical fingerprints of waste}. The three classes of chemical
fingerprints: Volatile Organic Compounds (VOC) (green), Persistent Organic Pollutants (POP) (orange), and HM (blue) left by the three BaC waste types Y1-Y18 (red), Y19-Y45 (purple), Y46-Y47 (yellow) in the top 28 countries at high risk of improper handling and disposal of wastes (HRIHDW).}
\label{alluvial}
\end{figure*}

The waste congestion of a given country can be produced either (i)
because it imports large amounts of waste from other countries, or
(ii) because it produces large amounts of waste which it cannot process
with its infrastructures. The first could be for instance the case
of China before 2017, where an estimate of $70\%$ of the world's e-waste ends
up in Guiyu, in Guangdong Province where no more than $25\%$ is recycled
in formal recycling centers. The second can be exemplified by the
case of household waste in Senegal, where the lack of infrastructures
and collection system makes the problem insurmountable by local authorities.
Senegal exported more than 15,000 tonnes of household waste to Italy
in 2009. To differentiate both situations we will designate them as
(i) congestion at arrival, for the case where congestion can be produced
by importing large amounts of a given type of waste; and (ii) congestion
at departure, for the case where congestion can be produced due to
the existence of large amounts of waste in a country, which are then
exported to another. Also, we should notice here that the amount of
waste of a given type reported by a country \texttt{A} as exported to a
country \texttt{B} is not always equal to the amount of the same waste reported
by \texttt{B} as imported from \texttt{A}. This difference could be due to several
causes which escape the analysis of the current work, but the split of congestion at arrival and at departure avoids any problem arising from this data asymmetry.

We then follow Lee et al. \cite{lee2019transient} and make a change of variable
in the model: $s_{i}\left(t\right)\coloneqq-\log\left(1-w_{i}\left(t\right)\right)$,
such that $s_{i}\left(t\right)$ represents the ``information content''
that country $i$ is not congested at time $t$.
Using this approach and considering $w_i\left(0\right)=w_{0}$ as an  initial condition, we define here the following models of waste congestion
in the W4 with dynamic memory:

i) Congestion at arrival
\begin{equation}
\begin{split}D_{t}^{\alpha}\, \mathbf{s}_{A}(t) & =\beta_{A}^{\alpha}\, A\, \mathbf{w}\left(t\right)\end{split}
,\label{eq:arrival}
\end{equation}

ii) Congestion at departure
\begin{equation}
\begin{split}D_{t}^{\alpha}\, \mathbf{s}_{D}(t) & =\beta_{D}^{\alpha}\, A^{T}\, \mathbf{w}\left(t\right)\end{split}
,\label{eq:departure}
\end{equation}
where $A^{T}$ is the transpose of the weighted adjacency matrix $A$ of the
network, and $\mathbf{s}_{A}$, $\mathbf{s}_{D}$ and $\mathbf{w}$ are vector formed by an element of $s$ and $w$ per node. 
Due to the uncertainties in the parameters involved in these equations
we consider here a ``worst-case-scenario'' approach in solving them.
That is, instead of obtaining the solution of these equations we will
solve a linearized version of them, whose solution is an upper bound to the exact solution \cite{abadias2020fractional}:
\begin{equation}
D_{t}^{\alpha}\mathbf{\hat{s}_{\ell}}(t)=\beta_{\ell}^{\alpha}\, \hat{B}\, \mathbf{\hat{s}}_{\ell}(t)+\beta_{\ell}^{\alpha}\, B\, \mathbf{b}\left(\mathbf{w_{0}}\right),\label{eq:upper}
\end{equation}
where $\mathbf{b}\left(\mathbf{w}_{0}\right)\coloneqq \mathbf{w}_{0}+\left(\mathbf{1}-\mathbf{w}_{0}\right)\log\left(\mathbf{1}-\mathbf{w}_{0}\right)$ with the logarithm taken entrywise, $\ell=\{A,D$\}, $\hat{B}=B\, \Omega$, $\Omega=\textnormal{diag}\left(\mathbf{1}-\mathbf{w}_{0}\right)$ ($\mathbf{1}$ is an all-ones vector), 
and $B=\left\{ A,A^{T}\right\} $. The solutions of these equations
(expressed as normalized amounts of waste, $\hat{\mathbf{w}}_{\ell}\left(t\right)$)
are nondivergent upper bounds to the exact solutions of the equations
(\ref{eq:arrival}) and (\ref{eq:departure}), respectively. That
is, $\hat{\mathbf{w}}_{\ell}\left(t\right)\succeq \mathbf{w}_{\ell}\left(t\right)$
where $\succeq$ indicates that the inequality is obeyed for every
entry of the vectors. Therefore, $\hat{\mathbf{w}}_{\ell}\left(t\right)$ represent
the worse-case-scenario of congestion at arrival and at departure
for every country in the W4. Let the initial condition be $\gamma = w_0 = 1-c/n$ with $c$ is a small number, i.e. $c \ll 1$, (we took $c = 0.005$), and $n$ is the number
of nodes in the network. 
Then the solution of (\ref{eq:upper}) is
given by (see SI for details) \cite{abadias2020fractional}:
\begin{align}
\mathbf{\hat{s}_{\ell}}(t) = & \left(\frac{1-\gamma}{\gamma}\right) \, E_{\alpha,1}\left(t^{\alpha}\, \beta_{\ell}^{\alpha}\, \gamma \,  B\right)\, \mathbf{1} \nonumber \\ & -\left(\frac{1-\gamma}{\gamma}+\log\gamma\right)\mathbf{1},\label{eq:final_model}
\end{align}
where $E_{\alpha,1}\left(t^{\alpha}\, \beta_{\ell}^{\alpha}\, \gamma \,  B\right)$
is the Mittag-Leffler matrix function of $B$ \cite{garrappa2018computing}.

The fact that the solution of (4) is given by means of Mittag-Leffler matrix functions implies that congestion parameters, e.g., congestion time, depends on the global interactions of countries in the network and cannot be derived from simple static (centrality) measures (see SI for explanation).
This model estimates the time at which a country saturates of a given
type of waste. From this value alone we cannot infer whether that
country is at risk of improper handling and disposal of wastes. To
illustrate this situation let us consider two countries which saturate
of type I waste at the same time: Japan and Afghanistan. They reach
their 50\% of congestion at $t=28654.54$ (risk of waste congestion
is 0.606). However, while Japan is one of the richest countries in
the world with GDP ranging 4.002-4.591 trillion USD, Afghanistan is
among the poorest ones with GDP of 7.521-21.972 billions USD for
the period of time considered here. This obviously gives these countries
very different capacities for managing a waste congestion, a situation
which is well reflected in the environmental track record of each
of these countries. We will account for these differences in the section
C. Potential Environmental Impact.

\subsection*{Structural analysis of W4}
During 2001-2019, the total amount of wastes reported by the BaC around
the world was of 1,470,096,618 metric tonnes (which is more than 4,000 times the weight of the Empire State building). Time-aggregated weighted-directed
networks of seven types of waste grouping together 108 BaC categories
were created as described in Methods. The distribution of wastes by
the different types considered here (see Methods) is very unequal
with a large concentration on the wastes of types I-III. These three
types of wastes account for 95.41\% of the total weight of wastes
traded in the period of study. We then focus here on these three types
and the rest are considered in the SI. 
Waste of type I accounts for 40.4\% of the total volume of wastes traded world-wide in the period of study, followed by type II (28.9\%) and type III (26.1\%).

For the period 2001-2019 most of the international trade of type I-III
wastes took place between developed nations. They accounted for 90.67\%
($8.29\times 10^7$ tons of type I), 70.19\% ($4.58\times 10^7$ tons of type II) and 99.07\% ($5.86\times 10^7$ tons of type III) of the total volume of waste traded in that period. 

A closer inspection of the W4 (see Fig. \ref{wastes percentage}) reveals a large imbalance
in the directionality of the HW trades between developed, developing
and least developed countries. In the case of wastes of type I--which
include clinical, medical, and pharmaceutical wastes as well as residues
from industrial waste disposal operations--developed nations exported
more to the developing and least developed world than what they import
from them, i.e., 4,340,000 and 25,500 tons, respectively. Even for
the case of household wastes (type III) developed nations exported
52,000 and 15,300 tons more than what they imported from developing
and least developed nations, respectively. Only in the case of wastes
of type II, which contains many valuable metals, the developed nations
imported more than what they exported to developing nations, i.e.,
9,870,000 tons. The exports and imports of types I-III display fat-tailed
distributions, indicating the existence of a relatively small number
of exporters/importers which concentrate most of the volume and number
of connections in the W4.

\begin{figure}
\begin{centering}
\includegraphics[width=0.46\textwidth]{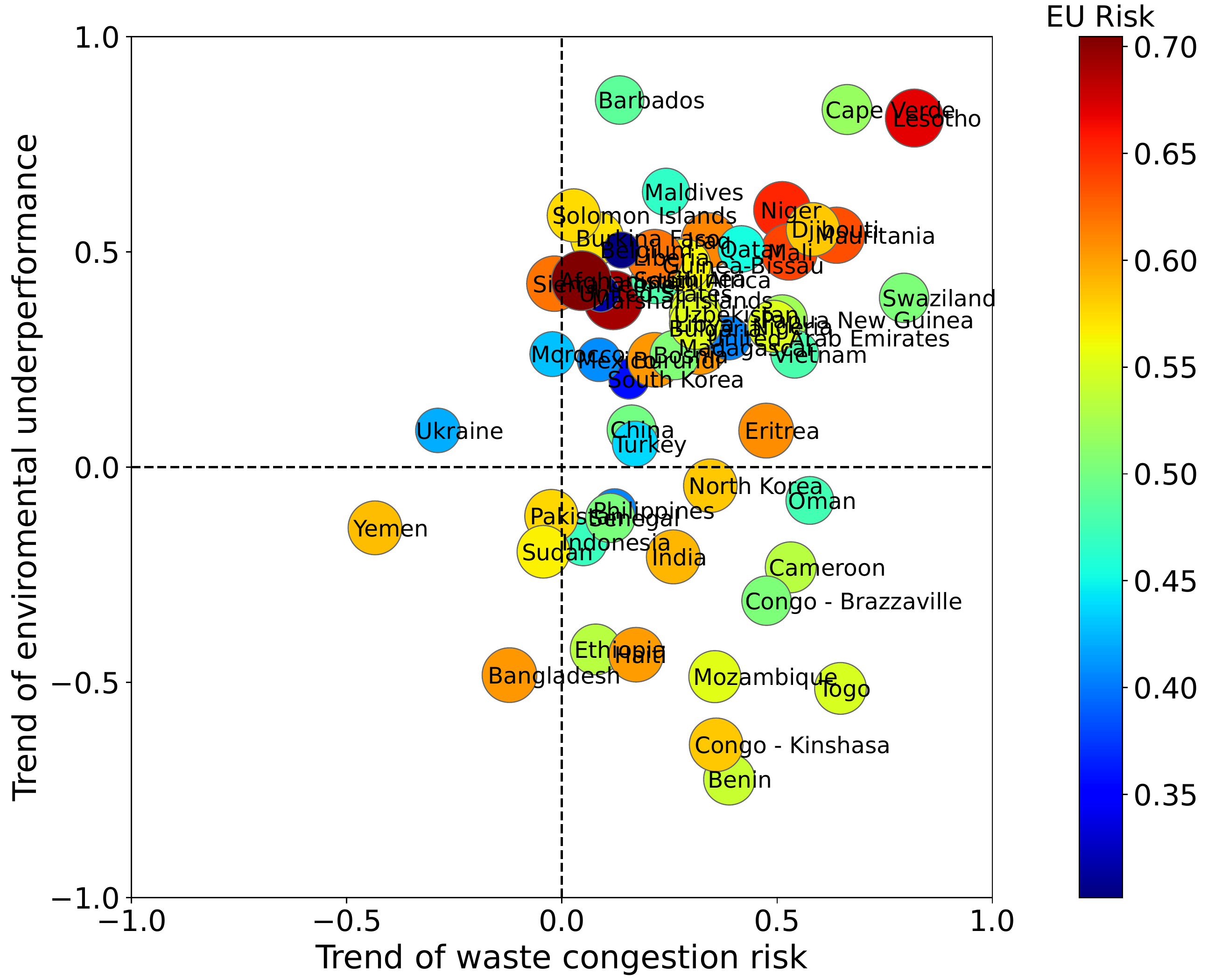}
\par\end{centering}
\caption{\textbf{Temporal trend (period 2001-2019) of the waste congestion risk and of the environmental underperformance risk for some countries at HRIHDW}.
The trend is measured by the Pearson correlation coefficient between
the corresponding variable and the years in the period. Bottom-left
quarter identifies the countries with a trend to improve both indices.
Top-right quarter identified those countries with a trend to deterioration
of both indices. EU Risk stands for Environmental Underperformance Risk.}
\label{temporal trends}
\end{figure}

\subsection*{Potential Environmental Impact}
As we have mentioned before, two countries with the same congestion
time of a given type of waste may have very different capacities for
processing it. These differences can be reflected in the 
environmental impacts that such waste have in the countries.
In order to capture these differences we use here the ``Environmental
Performance Index (EPI)'' measured by a set of parameters as described
in \cite{EPI}. Returning to the example of Japan and Afghanistan, which
saturate of waste type I at the same time, it should be noticed that
the EPI of Japan of 75.6 contrasts with that of Afghanistan of 29.5.
In  our  framework,  we  build  networks  by  considering HW type and  the  processing  capacities  of  the  countries  in  each waste  type.  Then we  calculate  the  corresponding  risk  that each  waste  poses  before  aggregating  to  a  global  score.
Thus, we introduce the Potential Environmental
Impact of Waste Congestion (PEIWC) (see Methods). In Fig. \ref{Potential Environmental Impact}(a),
we illustrate a typical PEIWC. Ideally, those countries with poor
EPI should manage low volumes of HW. They should appear at the top-left
corner of the PEIWC. Those countries with good EPI and low levels of
HW congestion should appear in the low-right corner of the PEIWC.
The central zone represents a ``tolerance'' zone, where countries
manage wastes according to their capacities and their environmental
responsibilities. However, there are countries with poor EPI that may
congest very quickly of waste. 
They are located over the tolerance zone 
and represent countries with high risk of improper handling
and disposal of wastes (HRIHDW).
\begin{figure}
\begin{centering}
\includegraphics[width=8.3cm]{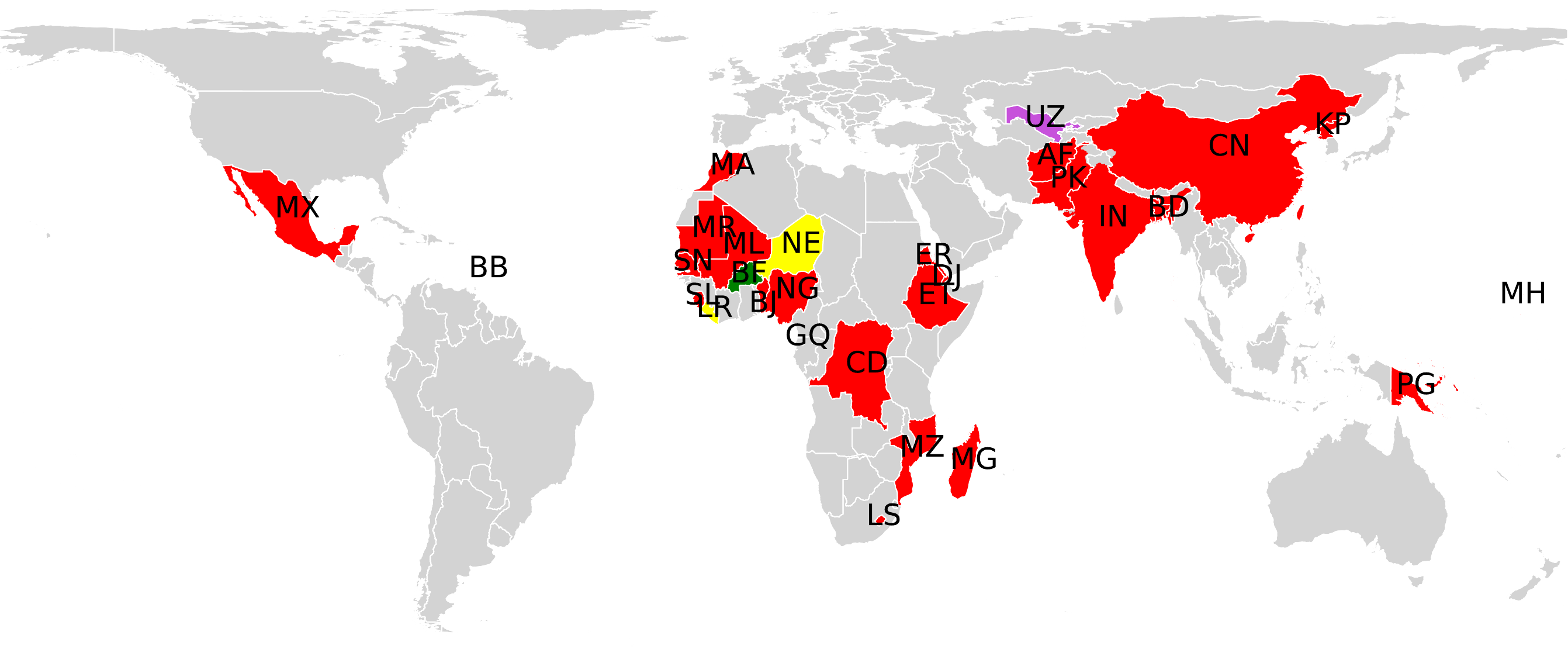}
\par\end{centering}
\caption{\textbf{High risk of improper handling and disposal of wastes}. Illustration of countries at HRIHDW of types I-III wastes and the chemical fingerprints left by these HW in their environment and/or human health. Countries with impact of heavy metals (HM) and persistent organic pollutants (POP) (green), volatile organic compounds (VOC) and HM (purple), VOC, HM and POP (red), VOC and POP (yellow) are illustrated. Map tiles by Bjorn Sandvik, under CC BY-SA 3.0 available at http://thematicmapping.org/downloads/world\_borders.php.}
\label{Other maps}
\end{figure}

In Fig. \ref{Potential Environmental Impact}(b-d), we illustrate
the PEIWC for wastes of types I-III. We identified 57 countries at
HRIHDW: 29 from Africa, 16 from Asia, 5 from the Americas, 4 from Europe and 3 from
Oceania. The color codes in Fig. \ref{Potential Environmental Impact} clearly reveal the geographical distribution of countries at different levels of risk of improper handling and disposal of wastes.

At the left-bottom side of the PEIWC there are a variety of countries/territories
of different sizes and levels of development, which are characterized
by processing efficiently the waste they receive, independently of
its amount. That is, these countries have developed capacities for
processing the amounts of waste they receive without compromising
the environment. This situation could be metastable and redirecting
more waste to these countries without increasing their processing
capacities could trigger waste-driven ecological problems.
However, due to the environmental good track record of these countries
they may represent opportunities for responsible investment in waste
treatment technologies.

It is difficult to find non-anecdotal evidence on the impact of waste trade on different countries in the W4. That is, if we want to move away from factual claims relying only on personal observations, collected in a casual or non-systematic manner about the impact of waste trade we should find some “markers” that can be traced quantitatively from waste to ecological impact. We propose here the use of Chemical Fingerprints (CF) as such markers for the analysis of the impact of waste trade on HRIHDW. A CF is a chemical or group of chemicals that are generated by wastes and leave a quantifiable trace on the environment. In the case of wastes of types I-III we have found that they may leave environmental and/or human health CF in the form of: (i) heavy metals (HM) \cite{alloway2013sources}, (ii) volatile organic compounds (VOC) \cite{he2019recent,niu2021temperature}, and persistent organic pollutants (POP) \cite{bogdal2013worldwide}. In Fig. \ref{alluvial} we illustrate the connections between the different kinds of wastes, the chemical fingerprints present in them and some of the countries found here at HRIHDW (see further discussion for a detailed analysis).

We also study waste-aggregated W4 networks for every year in the 2001-2019
period. Temporal trends of the waste congestion and environmental
underperformance risks were built for 57 countries at HRIHDW
(Fig. \ref{temporal trends}). What we plot here is the Pearson correlation coefficient of the waste congestion risk vs. time (x-axis) as well as of the EU vs. time (y-axis). Therefore, a negative value on the x-axis indicates that the corresponding
country has dropped its waste congestion risk. Similarly, a negative
value on the y-axis indicates that the country improved its environmental
performance.
Very few countries display a tendency to improve both risk indices (bottom-left quarter), while the majority showed simultaneous detriment of both (top-right quarter) from 2001 to 2019.
The values plotted here should not be confused with those in Fig.
\ref{Potential Environmental Impact}, where we plotted the absolute risks of waste congestion and of
environmental underperformance for countries. For instance, Lesotho
and Bangladesh are both at HRIHDW for waste of types I and II, i.e.,
they are at the top-right corner of Fig. \ref{Potential Environmental Impact} (b) and (c). However, Lesotho
is one of the countries which have worsened both, its waste congestion
risk and its environmental performance in the period 2001-2019 (top-right
corner of Fig. \ref{temporal trends}), while Bangladesh has significantly improved its
environmental performance and slightly improved its risk of waste
congestion (bottom-left corner of Fig. \ref{temporal trends}).

For a better characterization of the global structures of these 18
networks (one per year of the period analyzed), we calculated several of their topological features. The
edge density of these networks displays a significant decreasing tendency
along the period of time analyzed, i.e., the Pearson correlation coefficient
with time is $r=-0.69$. The trade networks have become ``smaller-worlds''
from 2001 to 2019 with a positive trend of the average
clustering coefficient ($r=0.59$) and a negative one of the average
path length ($r=-0.36$). The average reciprocity of edges between
pairs of connected countries was relatively stable across the period
($r=-0.05$). We should notice that this index does not account for the weights of the edges. However, the average number of weighted triangles displayed a
very strong negative tendency from 2001 to 2019 ($r=-0.83$), followed
by a similar trend of the weighted subgraph centrality of the countries
($r=-0.78$). The implications of these findings for the international
waste trade are analyzed in the next section.

\section*{Discussion}
Here we focus the discussion of the results found in the previous section on the analysis of CF in HRIHDW. For that purpose we performed an intensive literature search that allowed to trace back some CF generated by the types of wastes that produce the high risk in those countries to their environmental and/or animal/human health effect. As mentioned before, in Fig. \ref{alluvial} we illustrate the connections between the BaC wastes Y1-Y47, their CFs and the top 28 countries at HRIHDW (see Methods).
\begin{figure}
\begin{centering}
\includegraphics[width=0.5\textwidth]{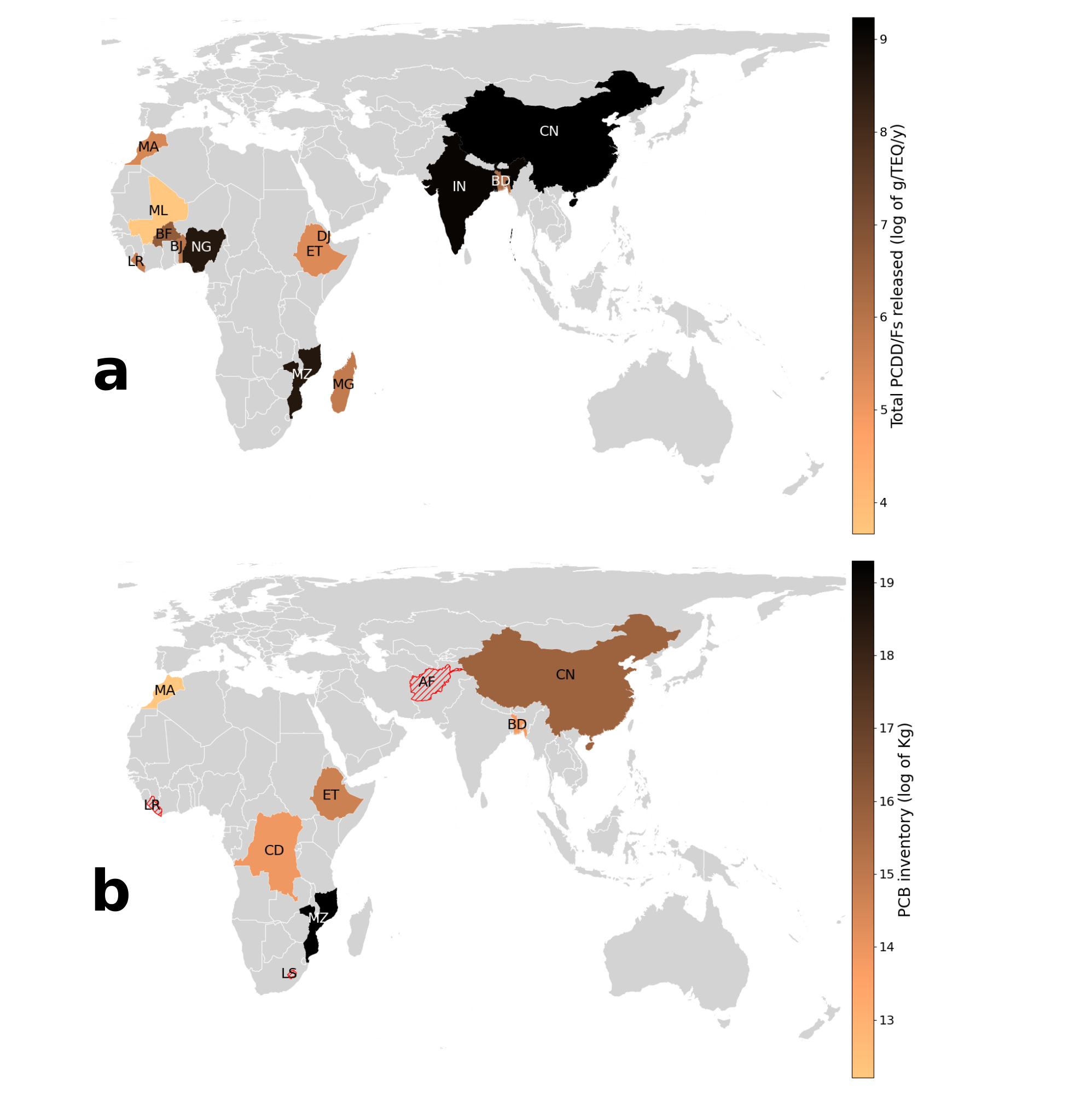}
\par\end{centering}
\caption{\textbf{Geographical distribution of PCB and Dioxins}. \textbf{a} Amounts of PCB stored in some of the countries at HRIHDW identified in this work. The amounts are given in logarithmic scale. \textbf{b} Total amounts of PCDD/Fs released to the environment by some of the
countries at HRIHDW identified in this work. The amounts are given
in logarithmic scale. The average amount of PCDD/Fs released in the
64 countries not in the list of countries at HRIHDW is 398.8 g/TEQ/y,
which in log scale is 5.99. Map tiles by Bjorn Sandvik, under CC BY-SA 3.0 available at http://thematicmapping.org/downloads/world\_borders.php.}
\label{PCB_Dioxins}
\end{figure}

\subsection*{Heavy metals}
Waste is one of the main anthropogenic sources of HM in the environment
\cite{alloway2013sources,ishchenko2018environment}, with electrical and electronic waste
(e-waste) alone containing 56 metals \cite{purchase2020global}.We
focus here on 8 HM ubiquitous in wastes of different kinds. Lead (Pb),
cadmium (Cd), nickel (Ni), mercury (Hg), chromium (Cr), zinc (Zn),
copper (Cu) and arsenic (As), appear in waste from pesticides, medicines,
paints, dyes, catalysts, batteries, electronic devices, industrial
sludge, printing products, incineration of household wastes, among
others \cite{alloway2013sources,ishchenko2018environment} (see SI).

In total, from the 28 countries at HRIHDW there are 24 ones in which
waste traded through the W4 can leave CF in the form of HM (see Fig. \ref{Other maps}). We have
found that the major sources of waste-generated HM are the open (unregulated)
dumpsites existing in many developing countries as well as the (informal)
recycling of wastes, principally of e-waste. To have a clearer picture
of the situation we refer to the data of the period 1990-2015 quantifying
the amount of hazardous waste landfilled and incinerated in some of
the countries found here at HRIHDW \cite{akpan2020hazardous}. In Niger,
Morocco and Madagascar, 1,057,000, 58,810 and 33,812 tons, respectively,
were landfilled, while 12,145 and 1,698 tons, were incinerated
in Madagascar and in Benin, respectively. Unregulated waste dumping
sites have been identified as the source of Pb, Hg, Ni, Cu, Cr, Cd
and Zn in the main source of ground water abstraction in southwestern
Burkina Faso, where informal settlements and peri-urban agriculture
place the population at risk \cite{sako2020hydrogeochemical}. Waste disposal
and its incineration have been identified in Kinshasa, Democratic
Republic Congo, among the main sources of Cd, Pb and Ni in ambient
air \cite{kabamba2016toxic}. Both illegal and legal waste dumping sites, among
other sources, have caused that HM such as Zn, As and Pb are significantly
present in sediment of Maqalika Reservoir, Lesotho. In particular,
As and Pb were found in common carp fish at concentrations higher
than the World Health Organization (WHO) permissible limits recommended for fish consumption,
placing the residents at significant health risks from the intake
of individual metals through fish consumption \cite{gwimbi2020heavy}.

Human and environmental damages produced by waste recycling are even
more dramatic. In the case of e-waste, in most developing countries
it is disposed of in domestic landfill sites and recycled in an informal
way. This typically involves burning materials for recovering copper,
and acid extraction to recover precious metals. Such practices are
common in China, India, Pakistan, and Nigeria \cite{ikhlayel2018integrated}, which
are all identified here as countries at HRIHDW. As a consequence,
in China \cite{song2015review}, levels of Pb in mother-infant
pairs were found to be five times higher in regions known for the
high concentration of e-waste disposal/processing than in control.
It was associated to the higher rates of adverse birth outcomes observed
in Guiyu--where 70\% of global e-waste ends up \cite{li2020environmental}--related
to control. In the same region children are reported to have significantly
higher levels of Pb, Cr, and Ni, which have been linked to low mean intelligence coefficient (IQ), and decreased forced vital capacity \cite{song2015review}.
Cd, Pb, Zn,Cu, Ni, As, and Cr were also found at higher levels in
hairs of residents and dismantling workers in Longtang and Taizhou
relative to control locations \cite{song2015review}. In India,
it has been reported that the levels of dermal exposure of HM in workers
of Indian e-waste recycling sites is 192.6 (Cr), 78.1 (Cu), 30.9 (Pb)
and 37.3 (Zn) times higher than those for people not exposed to e-waste \cite{singh2018health}. 

In Nigeria, on average, 400,000 second-hand
or scrap computers enter into the country annually, which represents
about 60,000 metric tonnes per annum \cite{nnorom2008electronic}. It is
estimated that the country generates 1,100,000 tons of e-waste \cite{akpan2020hazardous}.
Several reports have quantified high levels of HM contamination of
soils in e-waste dumpsites and of informal e-waste recycling in Nigeria
\cite{jiang2019impacts,isimekhai2017heavy,adeyi2017heavy}. Another waste-source of HM
is the dismantling of used electric batteries, mainly from cars. Madagascar,
which is one of the countries identified here at HRIHDW, is considered
as a regular destination for traffic of batteries, where Pb is extracted
by local scavengers and then send to China, Pakistan or Dubai \cite{cholez2020mundane}.
In Senegal, another country at HRIHDW, the death of 18 children \cite{haefliger2009mass} has been linked to high levels of Pb in children living in surrounding
areas used for recycling of used lead-acid batteries. In some cases
it is difficult to trace the HM to a particular country due to the
transborder nature of the region contaminated. This is the case of
the countries in the Gulf of Guinea--Benin, Cameroon, Ivory Cost,
Ghana, Nigeria--where contamination by toxic waste dumping is known,
which include high levels of heavy metals proceeding from e-waste
\cite{okafor2020toxic,scheren2002environmental}.

In Mexico, which is at HRIHDW, it has been
reported 8 abandoned or illegal hazardous waste sites in Baja California
and 15 ones in Coahuila, mainly containing HM. An estimated 6,000
tons of Pb wastes, as well as other HM including Sb, As, Cd and Cu
resulting from the battery recycling operation have been reported
in Tijuana, Baja California, Mexico \cite{jacott2001generation}. In Djibouti, contamination
of soils by As and Cr at concentrations 10 times higher than the US
maximum contaminant levels where produced by a shipment of containers
with up to 20 metric tons of chromate copper arsenate intended for
treating electric poles, which were found leaking in the port of Djibouti
\cite{kinaSUR}. In other countries found here at HRIHDW such as
Papua New Guinea, Uzbekistan and Bangladesh there are reports of HM
pollution affecting the environment and public health. However, it
is difficult to trace these polluting HM to hazardous wastes due to
the high impact of mining \cite{mudd2020mining,kodirov2018trace} as well
as industrial pollution \cite{kholikulov2021effect,ahmed2016human} in
these countries.

\subsection*{Volatile Organic Compounds}
VOC are ubiquitous organic pollutants affecting atmospheric chemistry
and human health \cite{he2019recent}. VOC can be released from wastes
containing solvents, paints, cleaners, degreasers, refrigerants, dyes,
varnishes and household wastes, from processing of e-wastes, plastics
and waste incineration \cite{he2019recent,niu2021temperature}. We
identify benzene (B), toluene (T), ethylbenzene (E) and o-, m-, and
p-xylenes (X) as potential CF of Y1-Y47 waste \cite{chen2021volatile,he2019recent,niu2021temperature,sabel1984volatile}.
Toluene is the only BTEX which has significant non-traffic sources,
with important contributions from previously mentioned sources. Indeed,
when the T/B ratio is over two it indicates the existence of waste sources
beyond vehicular traffic \cite{gelencser1997toluene}.

It is difficult to disentangle the possible sources for T/B ratios
in the countries at HRIHDW to retain only the information concerning
waste-generated VOC. For instance, in many developing countries T/B
values can be affected by unofficial gasoline selling places, or by
combustion processes for cooking in indoor kitchens. We have then
searched for some potential sources of large T/B ratios that can be
more likely be assigned to waste accumulation or processing. For instance,
several VOC have been identified in an e-waste dismantling town in
Guangdong province of China, including alkanes, BTEX, and organohalogen \cite{chen2021volatile}. The T/B ratio found here was 3.15, which clearly correlates
with emissions of VOC occurring during pyrolisis of e-waste \cite{chen2021volatile}.
T/B ratio of 9.36 is reported for Guangzhou \cite{chan2006characteristics}, which is the
capital city of Guangdong. In the city of Dakar, Senegal, both at
the urban district and at a semirural district, T/B ratios were 4.51
and 5.32 \cite{ba2019individual}. Senegal is a country at HRIHDW for types I, II
and III. In Senegal there has been continuous problems with the collection
of household waste \cite{kapepula2007multiple}, which have been responsible for public
health problems (dermatosis, diarrhea, conjunctivitis and malaria) \cite{thiam2017prevalence}.
Other HRIHDW countries with high values of T/B ratio reported at different
locations are the following: Bangladesh (6.85), Benin (7.75), Burkina
Faso (2.32), Ethiopia (2.3, 4.25), India (3.58, 3.67, 6.66, 8.97),
Mexico (2.19, 5.70, 6.59), (see SI for references).

\subsection*{Persistent Organic Pollutants}
POP are chemicals with high resistance to degradation in the environment,
high accumulation in human/ animal tissues and transmission through
food chains \cite{bogdal2013worldwide}. As POP indicators we consider here polychlorinated
biphenyls (PCB) \cite{liu2020critical} and polychlorinated dibenzo-p-dioxins and
polychlorinated dibenzofurans (PCDD/Fs) \cite{kanan2018dioxins}.

PCB are intentionally produced due to their many industrial applications.
They are related to neurodevelopment effects in infants, cancer and
immunotoxic effects in humans \cite{liu2020critical}. Vast amounts of PCB are stored
in some of the countries at HRIHDW (Fig. \ref{PCB_Dioxins} (a)) \cite{pen2016polychlorinated}. 
For instance, in Mozambique 240,571 tonnes of oil suspected to have
PCB are reported. Pollution by particulate and vapor samples containing
PCB was detected in three sites in KwaZulu-Natal Province, South Africa \cite{batterman2009pcbs}, which is close to the Mozambique border. PCBs are also found
in four fish species from Lake Koka, Ethiopia \cite{deribe2011bioaccumulation}, where 2,505
PCB-containing transformers with 1,182 tons of PCB oil and 40 PCB-containing
capacitors with 1.255 tons of PCB oil are reported. In China high
PCB concentrations have been reported in sediments from Pearl River
and its estuary \cite{xing2005spatial,cai2008status}. In Dalian Bay and Songhua River the
pollution by PCB is directly related to PCB equipment storage locations \cite{xing2005spatial}. In the Bengal coast of Bangladesh PCB contamination is linked
to the past and on-going use of PCB-containing equipment \cite{habibullah2019occurrence}.
Indeed, all 209 congeners of PCBs were found in 48 seafood samples
collected from the coastal area of Bangladesh, with severe health
risk for coastal residents \cite{habibullah2019polychlorinated}. In Bangladesh it is known to
exist 55.8 tons of PCB in use, 403 tons of contaminated oil contained
in waste equipment, 519 tons of contaminated waste transformer oils
and 22.5 tons of PCB contained in materials of old ships. Other three
countries found here at HRIHDW--Lesotho, Liberia and Morocco--have
large amounts of equipments (transformers and condensers) containing
PCB. In Morocco, for instance, 3,500 tons of oil with more than 50 mg/kg
of PCB are stored. There are 20 identured sites storing PCB wastes,
from which 50\% have been found to present floor pollution. Some African
countries identified here at HRIHDW have also large storages of PCB
waste. Nigeria has 341 transformers containing PCB. D. R. Congo has
188 PCB electrical transformers with 457 tons of PCB oils, 130,000
liters of pure PCB and 340,000 liters of PCB containing oil and in
Sierra Leone there are 103,372 tonnes of oil having PCB. The effects
of PCB pollution in these countries have been documented. For instance,
high levels of PCB has been found about 400 km off parts of the West
African coast \cite{gioia2011evidence}. The total amount of PCB in the serum of recent
immigrants who came from Sub-Saharan countries to the Canary Islands
(Spain) identified levels of 78, 124, 141, 181, 181, 255 ng/g lipid
for immigrants from Senegal, Guinea, Mali, Sierra Leone, D. R. Congo,
and Nigeria, respectively \cite{luzardo2014socioeconomic}. Importantly, the
authors of that report found that the immigrants' PCB levels were
strongly associated with the imports of second-hand e-waste by their
country of origin, supporting our hypothesis connecting waste to CF
and these to environmental/human damage.

On the other hand, PCDD/Fs are known to be extremely toxic in animals/humans \cite{kanan2018dioxins}. Consequently, their release to the environment are presented
as toxic equivalent (TEQ) (see Fig. \ref{PCB_Dioxins} (b)) \cite{fiedler2015release}. In
D. R. Congo alone PCDD/Fs amount to 300,412 g/TEQ/a (grams per toxic
equivalent per year) \cite{pius2019monitoring}. It is followed by China (10,232), India
(8,658), Nigeria (5,340), Lesotho (1,708) and Sierra Leone (1,242).
The mean TEQ of PCDD/Fs in 75 countries, excluding those found here
at HRIHDW, is 586.84 while those at HRIHDW is 2161.96 \cite{fiedler2015release}.

\subsection*{Evolution of the W4 in the period 2001-2019}

As we have seen the W4 has become slightly less densely connected
from 2001 to 2019. The most dramatic change, however, has been registered
by the drop of the average number of weighted triangles in the networks.
This index has dropped an order of magnitude from 2002 to 2019. Here
we are talking about directed triangles, that is, those in which a
directed path \texttt{A}$\rightarrow$\texttt{B}$\rightarrow$\texttt{C}$\rightarrow$\texttt{A} exists. Therefore, the number of directed
triangles can decrease due to (i) the deletion of some of the edges
forming the triangle, or (ii) due to the inversion of the direction
of any of the three arrows of the triangle. The fact that we observe
a positive trend in the clustering coefficient and a drop in edge
density along the period, incline us to think more about the second
possibility. In this scenario there are three equivalent possibilities:
(a) \texttt{A}$\leftarrow$\texttt{B}$\rightarrow$\texttt{C}$\rightarrow$\texttt{A}, (b) \texttt{A}$\rightarrow$\texttt{B}$\rightarrow$\texttt{C}$\leftarrow$\texttt{A}, and (c) \texttt{A}$\rightarrow$\texttt{B}$\leftarrow$\texttt{C}$\rightarrow$\texttt{A}.
In the original triangle every country \texttt{A}, \texttt{B}, and \texttt{C}, has one input
and one output, which make the system ``balanced'', but in (a),
(b) and (c), such balance is broken. In (a) the country \texttt{A} is becoming
a net importer, while \texttt{B} is a net exporter. The same scenario is repeated
for other nodes in (b) and (c). Although we focus here only on triangles
the situation is repeated for other cycles as revealed by the fact
that the weighted subgraph centrality of the countries also decay
with time in this period.

The countries which displayed the most negative trend in the number
of weighted triangles in the period 2001-2019 were: Belgium, Germany,
Netherlands, Spain, and France, closely followed by Ukraine, Canada,
Ireland and U.S. In order to investigate whether these countries have
evolved to net importers or to net exporters we use the difference
$\triangle \textbf{S}=\textbf{S}_{in}-\textbf{S}_{out}$ of the in- $\textbf{S}_{in}$ and out-strengths
$\mathbf{S}_{out}$ of every node in the networks. The in-strength accounts
for the weighted amounts of waste imported by a country, and the out-strength
for those exported. We found that Germany, France, U.S. and Ukraine
evolved from more balanced situations to become mainly net exporters
in this period, with Pearson correlation coefficients $r$ of $\triangle S$
vs. time of -0.78, -0.72, -0.66, and -0.44, respectively. Countries
like Netherlands ($r\approx0.88$), Belgium ($r\approx0.57$), Spain
($r\approx0.42$) and Canada ($r\approx0.34$) evolved into net importers
in the period.

In the general framework of the W4 evolution in the period 2011-2019,
the countries displaying a more significant transformation to net
exporters are Slovenia, U.K., New Zealand, and Germany, followed by
France, and U.S. at different places among the top 15. On the other
side, those transforming into net importers, the list is headed by
Netherlands, Poland, Sweden and R. Korea. Among the countries at HRIHDW
the main transformation towards net exporter is observed in China
($r\approx-0.70$) and those becoming major importers (in terms of the volume of waste) are Mexico ($r\approx0.66$),
India ($r\approx0.62$) and Uzbekistan ($r\approx0.47$). In 15 of
the countries at HRIHDW we found that $S_{in}=0$ for every year
reported. That is, these countries did not reported to have imported
any waste during 2001-2019 but they export significant amounts of
waste as to place them at HRIHDW. This situation in some of these
countries could be a red alert of illegal waste imports which of course
are not reported at the Basel documents (see Fig. 4 in \cite{akpan2020hazardous}). 

We also considered the betweenness centrality $BC$ of the countries
in the W4. The largest BC is observed in developed countries like U.K., France,
Germany, Austria, Netherlands, and Belgium, while 116 countries, mainly
developing ones, have BC equal to zero. We identified that
82.8\% of the countries at HRIHDW have zero betweenness. That is,
they are endpoints (net exporters or net importers) in the W4, while
the developed countries have a more balanced situation in which imports
and exports occur simultaneously.

We have introduced here a mathematical framework that allows to model
the flows of waste through the international network of waste trade.
This model allows to identify the time at which a given country reaches
its carrying capacity and consequently it is considered to be congested
or saturated of a given type of waste. Although we have merged several
classes of waste into groups to facilitate the modeling and interpretation
of results, the model can be adapted to individual waste categories
according to BaC. Using this strategy we have identified countries
which are at HRIHDW due to their relatively fast congestion of waste
and their poor track record on environmental performance. The causes
producing such potential waste saturation can be multiple and are
not explored in this work. However, the current results trigger some
red alert about the critical situation of some countries and the necessity
of substantial investment in waste management at a global level. 

The theoretical model presented here as well as the main results of
this work can be applied to study the impact of different emerging
scenarios, such as:
(i) ``Import bans\textquotedblright{} policies in major importers,
like the one imposed in 2017 by China \cite{balkevicius2020fending}. Consider a country
$i$ that exports HW to the set of countries $\eta$. If $j\in\eta$
imposes a ban on HW, then the model can be used to redistributing
the amount of wastes exported by $i$ to a country $j$ among the
rest of countries in $\eta$. We can analyze how this redirectioning
impacts on the congestion time of the countries in $\eta$ and on
the rest of the World;

(ii) Understanding the potential waste congestion problems arising
from the COVID-19 pandemic and from emerging sources of e-waste \cite{liang2021repercussions,Heath2020_NatEnergy}. Consider the set of countries that export/import a given type
of waste, e.g., biomedical waste or e-waste. Increase those amounts
in the model according to reported data or estimations, and calculate
the congestion time for these countries as well as for the rest of
the World;

(iii) Analyzing the impact of a global scenario of increasing amounts
of all wastes. Use the model to proportionally increase the amounts
of waste of every country/territory and run the simulation for determining
the congestion time of every country, which can be compared with the
results previous to the increase of waste amounts.

\section*{Methods}

\subsection*{Data collection\label{sec:Data-quality-and}}
We extract the data used to build W4 networks from BaC Online Reporting
Database~\cite{Basel_1,Basel_2}. It contains summarized compendiums
where individual national reports are altogether condensed into single
Excel files per year, with the explicit and quantitative information
of associated parties: destination, import, origin and transit. We
extract from these files the information about countries/territories
of exports and imports, transaction amounts in metric tons,
waste classification codes, characteristics and type of waste streams.
Code names of countries,
and special territories like those that has no total political sovereignty,
are considered by using the standard ISO 3166-1 alpha-2 \cite{Countries}.
We do not include the countries of transit due to its scarcity in
the reports, and because of the lack of information about the temporary
order of the landings. We also excluded the existing self-export (a
country that exports to itself). We manually curated the database
for errors in the country/territories names, e.g., due to typos or
possible transcription errors, as well as for the use of nonofficial
country codes such as EIRE instead of IE for Ireland. The BaC reports
may also combine formal ISO alpha-2 codes with others codes that
have become obsolete and sometimes with codes of another standards
like transitory codes or international postal union codes. Reports
may pointing out to a state party that currently is dissolved or split
into two new ones, e.g., Serbia and Montenegro. In the case of waste
categories we also exclude those for which their codes do not coincide
with the ones defined by the BaC, such as 11b, AN8, Y48.

\subsection*{Waste types}

We consider 108 categories of wastes according to BaC classification,
which are then grouped into seven types of waste designated by Type
I-VII. The classification of wastes used in this work are based on
the Annexes I, II and VII of the BaC \cite{Basel_1}. No wastes in
categories B of the BaC are included in this work as they are not reported
by countries in the database of the Convention \cite{Basel_2}.

\textbf{Type I} considers, for instance, Y1: Clinical wastes from
medical care in hospitals, medical centers of clinics, Y2: Wastes
from the production and preparation of pharmaceutical products, up
to Y18: Residues arising from industrial waste disposal operations
(see pp 46 of Ref. \cite{Basel_1}). The \textbf{Type II} of wastes
used in this work associates the second subdivision of the Annex I,
Y-codes Y19-Y45. In general, wastes containing 27 chemical constituents,
i.e., Y19: Metal carbonyls, Y20: Beryllium compounds, up to Y45: Organohalogen
compounds. The \textbf{type III} of wastes discussed here accounts
for the Annex II of the BaC classification. Y46: Wastes collected from
households, and Y47: Residues arising from the incineration of household
wastes. A complete list is provided in the SI.

The remaining four types of wastes recover the four subclassification
of the Annex VIII \cite{Basel_1}. Specifically, \textbf{Type IV}
links with the Metal and Metal-Bearing Wastes. It accounts for A-list
items grouped from A1010-A1090 and A1100-A1190, e.g., A1010: Metal
wastes and waste consisting of alloys of Antimony, Arsenic, Cadmium,
Selenium, among others; up to A1190: Waste metal cables coated or
insulated with plastics containing or contaminated with coal tar,
PCB11, lead, cadmium, other organohalogen compounds (see pp 66 of
Ref. \cite{Basel_1}). \textbf{Type V} relates Inorganic constituents
containing metal and organic material. (cathode-ray glasses, liquid
inorganic fluorines, catalysts, gypsum, dust-fibres of asbestos, coal-fired
power plant fly-ash). Its A-items ranges from A2010-A2060. \textbf{Type
VI} associates Organic constituents containing metal and inorganic
material. (Petroleum coke and bitumen, mineral oils, leaded anti-knock
sludge, thermal fluids, resin, latex, plastiticizers, glues, adhesives,
nitrocellulose, phenols, ethers, leather wastes, (un)halogenated residues,
aliphatic halogenated hydrocarbons, vinyl chlorides), accounting for
A-items: A3010-A3090, A3100-A3190 and A3200. Finally, \textbf{Type
VII} are Wastes which may contain either inorganic or organic constituents
(Some pharmaceutical products, clinical-medical-nursing-dental-veterinary
wastes from patients and researches, biocides-phytopharmaceutical,
pesticides, herbicides outdated, wood chemicals, (in)organic cyanides,
oils-hydrocarbons-water mixtures, inks, dyes, pigments, paints, lacquers,
varnish, of explosive nature, industrial pollution control devices,
for cleaning of industrial off-gases, peroxides, outdated chemicals,
from research or teaching activities, spent activated carbon, to name
a few). It accounts for the groups A4010-A4090, and A4100-A4160.

\subsection*{W4 construction}

We construct a weighted directed network for each of the types
of waste analyzed. In every network the nodes correspond to the countries/territories
reporting the given type of waste in the period 2001-2019. It is frequent
in the BaC database that a country $i$ reports the export (import)
of an amount $q_{ij}$ to (from) $j,$ which includes several BaC waste
categories. If all the BaC categories belong to the same waste type, then we simply use that amount as the weight of the link $(i,j)$.
However, it happens sometimes these BaC categories belong to several
waste types. Let us consider two BaC categories $C_{1}$ and $C_{2},$ e.g.,
Y1 and Y19. Then, $C_{1}$ belongs to one waste type, e.g., type I,
and $C_{2}$ to another, e.g., type II. In this case we have to split
the quantity $q_{ij}$ in the weights of the links between $i$ and
$j$ for the two types of wastes. We then proceed as follows. We obtain
the weight of the link $(i,j)$ for the waste of type $k$ as
\begin{equation}
w_{ij}^{k}=\frac{q_{ij}\cdot\phi_{k}}{\Phi}\ ,
\end{equation}
where $\phi_{k}$ is the average of the amounts of waste of type $k$
traded between every pair of countries during the corresponding year,
and $\Phi=\sum_{k}\phi_{k}$ where the summation is carried out for
all types of waste involved in the quantity $q_{ij}$.

In any case we can obtain two different weights for a pair of countries
based on the data reported at BaC from ``Export'' and ``Import''
reports. Then we can have the following two different cases: (a) that
the amount $E\left(i,j\right)$ reported by country $i$ as exported
to country $j$ coincides with the amount $I\left(j,i\right)$ reported
by $j$ as imported from $i$; (b) that $E\left(i,j\right)\neq I\left(j,i\right)$.
In the case (a) we simply add a directed arc from $i$ to $j$ with
the weight $E\left(i,j\right)=I\left(j,i\right).$ In the case (b)
we assume that $i$ exports $\max\left[E\left(i,j\right),I\left(j,i\right)\right]$
to $j$. We designate by $\tilde{A}=\tilde{A}\left(G\right)$ the
adjacency matrix of the network $G.$ Notice that $\tilde{A}$ is
not necessarily symmetric because $\tilde{A}_{ij}=\max\left[E\left(i,j\right),I\left(j,i\right)\right]$
is not necessarily the same as $\tilde{A}_{ji}=\max\left[E\left(j,i\right),I\left(i,j\right)\right]$.
Here we normalize the adjacency matrices by: $A=\tilde{A}/\sum_{i,j}\tilde{A}_{ij}$.

\subsection*{Network parameters of the W4 networks}

Because the W4 networks are weighted and directed we consider here
the distributions of their in- and out-strengths, $\mathbf{S}_{in}$ and $\mathbf{S}_{out}$,
respectively. The in-strengths of the node $i$ is the sum of the
weights of all links pointing to $i$. The out-strength of that node
is the sum of the weights of all links leaving that node. For each
kind of strength we tested 17 types of distributions \cite{kotz1993continuous,kotz1994balakrishan}: beta,
Birnbaum- Saunders, exponential, extreme value, gamma, generalized
extreme value, generalized Pareto, inverse Gaussian, logistic, log-logistic,
lognormal, Nakagami, normal, Rayleigh, Rician, $t$-location-scale,
and Weibull. To test the goodness of fit we used \cite{maydeu2010goodness}: negative
of the log likelihood, Bayesian information criterion, Akaike information
criterion (AIC), and AIC with a correction for finite sample sizes.
The results are given in the Supplementary Information.

For the global characterization of the W4 networks we used the following
structural parameters (see \cite{estrada2012structure}):
\begin{enumerate}[label=(\roman*)]
\item Edge density, $\delta\left(G\right)$
\begin{equation}
\delta\left(G\right)\coloneqq\dfrac{m}{n\left(n-1\right)},
\end{equation}
where $n$ is the number of nodes and $m$ is the number of directed
edges;
\item Reciprocity, $\rho\left(G\right)$
\begin{equation}
\rho\left(G\right)\coloneqq\dfrac{r-\bar{a}}{1-\bar{a}},
\end{equation}
where the network is binarized before the calculation, $\bar{a}$
measures the ratio of observed to possible directed links and
\begin{equation}
r=\dfrac{L^{\leftrightarrow}}{m},
\end{equation}
with $L^{\leftrightarrow}$ being the number of reciprocal edges and
$m$ the total number of directed edges (see \cite{garlaschelli2004patterns} for
details);
\item Average number of weighted directed triangles, $\bar{t}$
\begin{equation}
\bar{t}\coloneqq\dfrac{tr\left(A^{3}\right)}{3n},
\end{equation}
where $tr\left(.\right)$ is the trace of the matrix $A$, and $A$
is the adjacency matrix of the network;
\item Average clustering coefficient, $\bar{C}$
\begin{equation}
\bar{C}\coloneqq\dfrac{1}{n}\sum_{i=1}^{n}\dfrac{2t_{i}}{k_{i}^{tot}\left(k_{i}^{tot}-1\right)-2k_{i}^{\leftrightarrow}},
\end{equation}
where $t_i$ is the number of directed triangles through node $i$, 
$k_{i}^{tot}$ is the sum of in- and out-degree of the node
$i$, and $k_{i}^{\leftrightarrow}$ is the reciprocal degree of $i$
(see \cite{fagiolo2007clustering} for details);
\item Average path length, $\bar{l}$
\begin{equation}
\bar{l}\coloneqq\dfrac{1}{n\left(n-1\right)}\sum_{i,j}l_{ij},
\end{equation}
where $l_{ij}$ is the length of the shortest path distance between
the nodes $i$ and $j$, where in case that there is not a path between
two nodes the value of zero is assigned to its length;
\item Average subgraph centrality, $\bar{S_{C}}$ 
\begin{equation}
\bar{S_{C}}\coloneqq\dfrac{1}{n}tr\left(e^{A}\right),
\end{equation}
where $\exp\left(A\right)$ is the matrix exponential of the adjacency
matrix.
\end{enumerate}

For the analysis of the individual countries of the network, apart
from the in- and out-strengths, we also analyze the following local
measures of W4 {[}79{]}. We refer the indices to a node labelled as
$i$:
\begin{enumerate}[label=(\roman*)]
\item Number of weighted directed triangles
\begin{equation}
t_{i}\coloneqq\left(A^{3}\right)_{ii};
\end{equation}
\item Subgraph centrality
\begin{equation}
SC_{i}\coloneqq\left(e^{A}\right)_{ii};
\end{equation}
\item Betweenness centrality
\begin{equation}
BC_{i}\coloneqq\sum_{p\neq i\neq q}\dfrac{\varrho_{piq}}{\varrho_{pq}},
\end{equation}
where $\varrho_{piq}$ is the number of weighted directed shortest
paths from $p$ to $q$ that goes through the node $i$ and $\varrho_{pq}$
is the total number of weighted directed shortest paths from $p$
to $q$.
\end{enumerate}

\subsection*{Congestion dynamics}

We consider the same rate for both, congestion at arrival and congestion
at departure, processes, i.e., $\beta_{A}=\beta_{D}$. In the simulations
we use $\alpha=0.75$, $\beta=0.01$, and $c=0.005$. As a way of
quantifying how easy a country get congested by a given waste we use
the time at which $50\%$ of the total congestion is reached, which is more accurate to determine that the time at which 100\% of saturation is reached. Let
us designate by $\hat{t}_{i}$ this time. Then, $\hat{t}_{i}$ is
the time $t$ at which $s_{\ell}\left(t\right)=0.5$.

We will illustrate the intuition behind the use of separated equations for congestion at arrival and at departure with an example. Let us consider a trade network of three countries where $A$ exports
100 tonnes of waste to $B$ and 120 tonnes to C; $B$ exports 200
tonnes to $C$, and $C$ exports 50 tonnes to $A$. 
Let the times $\hat{t}_{C}<\hat{t}_{A}<\hat{t}_{B}$ for the congestion at arrival model (see SI Fig.{[}8{]} for graphic illustration.

This indicates that $C$ is at
the highest risk of congestion due to its large imports of waste.
However, if we consider the process at departure, $\hat{t}_{A}<\hat{t}_{B}<\hat{t}_{C}$,
which indicates the highest risk at node $A$ due to the existence
of large amounts of this waste at the node.

\subsection*{Potential Environmental Impact of Waste Congestion}

We first define here the risk of waste congestion for a given country
as
\begin{equation}
R_{i}\coloneqq1-\hat{t}_{i}/\max_{j}\hat{t}_{j},
\end{equation}
where $i$ represents a given country, $\hat{t}_{i}$ is the congestion
time for the country $i$ either by importing or by exporting wastes
of a given type. That is, if $t_{1/2}\left(i\leftarrow\right)$ and
$t_{1/2}\left(i\rightarrow\right)$ are the times at which country
$i$ reaches 50\% of congestion by importing and exporting a given
type of waste, respectively, then $\hat{t}_{i}=\min\left[t_{1/2}\left(i\leftarrow\right),t_{1/2}\left(i\rightarrow\right)\right]$.
The index $R_{i}$ is normalized between zero (no risk) and one (maximum
risk) of congestion of wastes of a given type.

Due to the socio-economic differences between the countries in the
world, the use of $R_{i}$ along could be of little practical value.
For instance, for wastes of type I the Netherlands and Burkina Faso
have about the same value of $R_{i}$, which is near 0.99. For the same type of wastes Ireland and Ivory Cost also
have $R_{i}\approx0.89$. The situation is similar for waste of type
II, where the first pair of countries have $R_{i}=1$ and the second
pair have $R_{i}\approx0.94$. However, while Netherlands and Ireland
are among the richest countries in the world with GDP ranging 578-868
billions USD (Netherlands) and 164-236 billions USD (Ireland), the
other two countries are among the poorest with GDPs of 4.7-9.4 billions
USD (Burkina Faso) and 15-24 billions USD (Ivory Cost) for the
period of time considered here. This obviously gives these countries
very different capacities for managing a waste congestion, a situation
which is well reflected in the environmental track record of each
of these countries. The Environmental Performance Index (EPI), published
by the Universities of Yale and Columbia \cite{EPI}, quantifies the
performance of every country using sixteen indicators reflecting United
Nations\textquoteright{} Millennium Development Goals. They are accounted
for by six well-established policy categories (see Policymakers' Summary
at \cite{EPI}): Environmental Health, Air Quality, Water Resources,
Productive Natural Resources, Biodiversity and Habitat, and Sustainable
Energy, such that it covers the following two global goals: (1) reducing
environmental stresses on human health, and (2) promoting ecosystem
vitality and sound natural resource management. Then, while Netherlands
and Ireland are among the top environmental performers in the 2001-2019
period with average EPIs larger than 70 out of 100, Burkina Faso and
Ivory Cost are the bottom of the list with average EPIs of 45.2
and 55.9, respectively. We can account the risk of environmental underperformance
by an index bounded between zero and one as: $U_{i}=1-EPI\left(i\right)/100$.
PEIWC are defined by plotting the waste congestion risk $R_{i}$ for
a given type of waste versus $U_{i}$. For the demarcation of the
tolerance zone we use here the following. We obtain the linear regression
model that best fit $U_{i}$ as a linear function of $R_{i}$. Then,
the tolerance zone is defined by the upper and lower 50\% prediction
bounds for response values associated with this linear regression
trend between the two risk indices. The value of 50\% is used here
as a very conservative definition of the tolerance zone. Widening
this zone too much will make that almost no country is at HRIHDW,
which does not reflect the reality. On the contrary, narrowing it
to much will simply split countries into two classes, which will make
difficult to identify those at the highest risk of environmental underperformance
due to waste congestion.
Although we have identified 58 countries over this tolerance zone, i.e., the countries at HRIHDW, we performed our studied on a subset of them formed by 28 countries. These countries, referred in the text as "top 28", were selected by picking in every PEIWC those countries in the top 25\% of deviation over the tolerance zone. We then merged the sets of countries found at every PEIWC conforming this list of top 28 at HRIHDW.

\section*{Data availability}

Extracted set of Export and Import networks generated in this study is  available in 
(https://github.com/JohannHM/Fractional-congestion-Dynamics). J.H. Mart\'inez, S. Romero, J.J. Ramasco, E. Estrada, The world-wide waste web. JohannHM/Fractional-congestion-Dynamics: The World-wide waste web. Data and code (v1.0.0). Zenodo. https://doi.org/10.5281/zenodo.5786874 . (2021).
All raw data of the manuscript and its Supplementary Information was obtained directly from the Basel Convention web page: \hash{http://www.basel.int/Countries/NationalReporting/ElectronicReportingSystem/tabid/3356/Default.aspx}.
\section*{Code availability}
Custom MATLAB code is available on GitHub
(https://github.com/JohannHM/Fractional-congestion-Dynamics). J.H. Mart\'inez, S. Romero, J.J. Ramasco, E. Estrada, The world-wide waste web. JohannHM/Fractional-congestion-Dynamics: The World-wide waste web. Data and code (v1.0.0). Zenodo. https://doi.org/10.5281/zenodo.5786874 . (2021).


\begin{acknowledgments}
The authors thank J. Gómez-Gardeñes, M. Chavez, and A. Martínez for important suggestions. Partial funding has been received from the Maria de Maeztu program for Units of Excellence MDM-2017-0711 (J.H.M., S.R., J.J.R., and E.E.), from the grant PID2019-107603GB-I00 (E.E.) of the MCIN/AEI/10.13039/501100011033/, from the grant PACSS RTI2018-093732-B-C22 (J.J.R.) of the MCIN/AEI /10.13039/501100011033/ and of EU through FEDER funds (A way to make Europe), and also from the Colciencias Call (\#811) (J.H.M.) of the Colombian Ministry of Science, Technology and Innovation.
\end{acknowledgments}

\section*{Author Contributions}
E.E. designed, directed and wrote the manuscript. J.H.M., and S.R. contributed with extraction and curation of data, performed simulations and draw figures. J.H.M, J.J.R, and E.E. performed  computations, analyzed the results, and revised the manuscript.
\section*{Competing interests}
The authors declare no competing interests.
\section*{Additional information}
Supplementary Information is available for this paper.


\end{document}


\title{Supporting Information for \\
"The world-wide waste web"}

\author{Johann H. Mart\'inez, Sergi Romero, Jos\'e J. Ramasco and Ernesto Estrada}

\date{}
\maketitle

This material contains relevant innformation about supplementary dicussion and supplementary mehtods related to the connectivity distribution of the w4 and different types of network centralities; supplementary data, tables and figures with specifications of the chemical fingerprints, as well as for the waste classificacion.

\section*{Supplementary information guide}

\begin{itemize}
  \item Degree distributions of W4 networks.
  \item Chemical fingerprints common in hazardous waste.
  \item Centrality of countries in the W4 networks.
  \item Susceptible-waste congested model.
  \item Chemical fingerprints.
  \item Countries/territories without EPI.
  \item Waste categories in types IV-VII.
  PEIWS analysis of wastes types IV-VII.
  
\end{itemize}

\section*{Degree distributions of W4 networks}

For each of the type I-III we calculated the in- and out-strengths
(weighted degrees) of its nodes and tested 17 probability distribution functions:
beta, Birnbaum-Saunders, exponential, extreme value, gamma, generalized
extreme value, generalized Pareto, inverse Gaussian, logistic, log-logistic,
lognormal, Nakagami, normal, Rayleigh, Rician, t-location-scale, and
Weibull. The goodness of fit is tested by calculating the following
parameters: negative of the log likelihood (NlogL), Bayesian information
criterion (BIC), Akaike information criterion (AIC), and AIC with
a correction for finite sample sizes (AICc). The results are as follows.

\subsection*{Waste type I}

\renewcommand{\thetable}{SI. \arabic{table}}

\begin{table}[H]
\begin{centering}
\begin{tabular}{|c|c|c|c|c|c|}
\hline 
No. & distribution & NlogL & BIC & AIC & AICc\tabularnewline
\hline 
\hline 
1 & generalized pareto & -3249.47 & -6483.38 & -6492.94 & -6492.80\tabularnewline
\hline 
2 & generalized extreme value & -2196.84 & -4378.12 & -4387.68 & -4387.54\tabularnewline
\hline 
3 & t-location scale & -1548.53 & -3081.49 & -3091.05 & -3090.91\tabularnewline
\hline 
4 & exponential & -749.54 & -1493.90 & -1497.08 & -1497.06\tabularnewline
\hline 
5 & logistic & -535.26 & -1060.14 & -1066.51 & -1066.45\tabularnewline
\hline 
6 & normal & -371.55 & -732.72 & -739.10 & -739.03\tabularnewline
\hline 
7 & extreme value & -206.90 & -403.42 & -409.80 & -409.73\tabularnewline
\hline 
\end{tabular}
\par\end{centering}
\caption{Values of the statistical parameters quantifying the goodness of fit
of the distributions fitting the in-strength of the nodes of the W4
network of wastes type I. Only the top seven distributions are shown.}
\end{table}

\begin{table}[H]
\begin{centering}
\begin{tabular}{|c|c|c|c|c|c|}
\hline 
No. & distribution & NlogL & BIC & AIC & AICc\tabularnewline
\hline 
\hline 
1 & generalized extreme value & -1298.60 & -2581.63 & -2591.20 & -2591.06\tabularnewline
\hline 
2 & generalized pareto & -1291.81 & -2568.05 & -2577.62 & -2577.48\tabularnewline
\hline 
3 & t-location scale & -1166.20 & -2316.85 & -2326.41 & -2326.27\tabularnewline
\hline 
4 & beta & -1012.85 & -2015.33 & -2021.70 & -2021.63\tabularnewline
\hline 
5 & exponential & -749.54 & -1493.90 & -1497.08 & -1497.06\tabularnewline
\hline 
6 & logistic & -542.04 & -1073.71 & -1080.09 & -1080.02\tabularnewline
\hline 
7 & normal & -433.29 & -856.20 & -862.57 & -862.50\tabularnewline
\hline 
\end{tabular}
\par\end{centering}
\caption{Values of the statistical parameters quantifying the goodness of fit
of the distributions fitting the out-strength of the nodes of the
W4 network of wastes type I. Only the top seven distributions are
shown.}

\end{table}

\renewcommand{\thefigure}{SI. \arabic{figure}}

\begin{figure}
\begin{centering}
\includegraphics[width=1\textwidth]{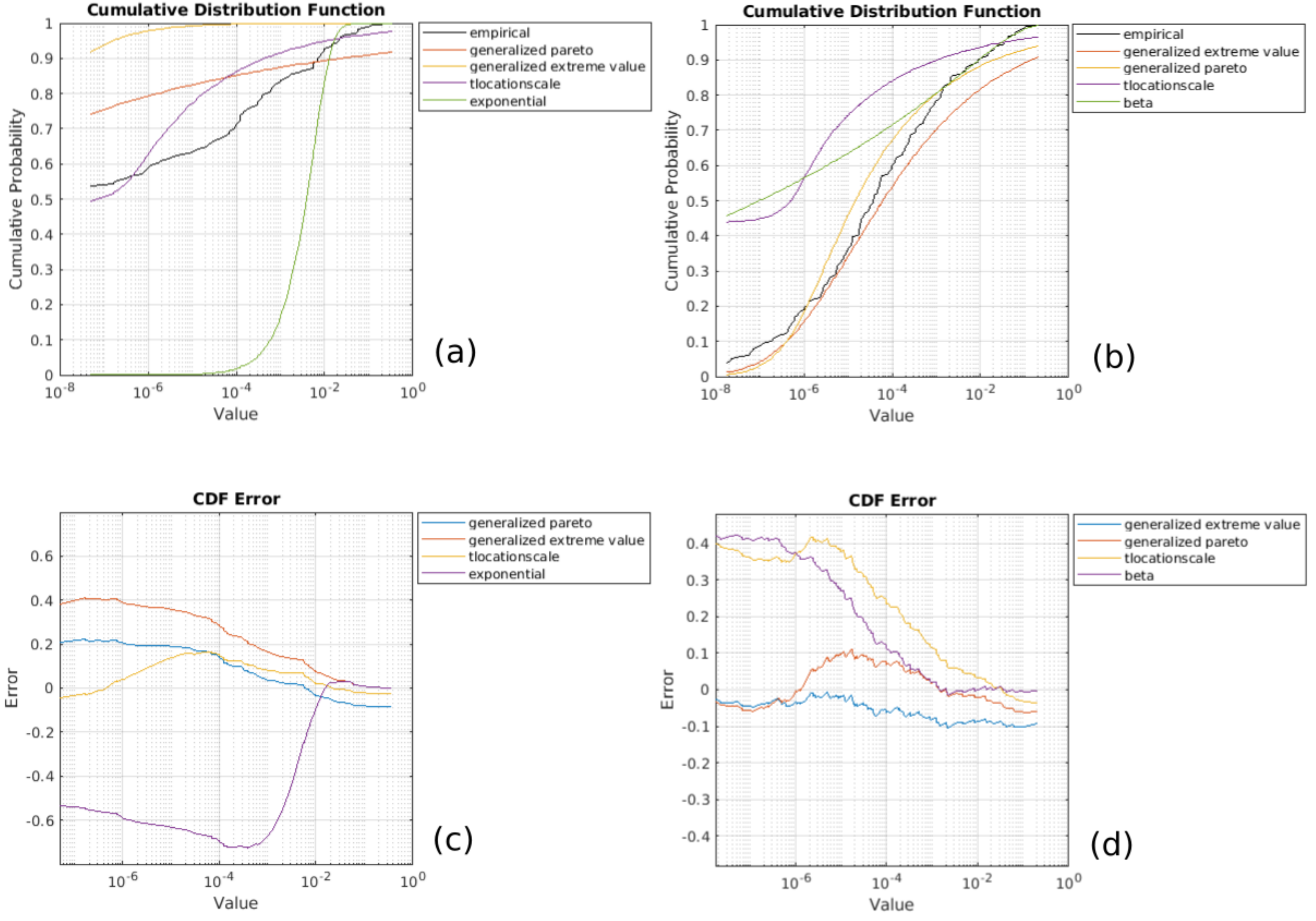}
\par\end{centering}
\caption{Cumulative in- (a) and out-degree (b) distributions and the corresponding errors (c)-(d) for the W4 of type I
wastes. The empirical distribution refers to the data of the W4 network
and the others correspond to the best fits using different kinds of
distributions, e.g., lognormal, loglogistic, generalized extreme value,
etc. The errors are obtained as the differences of the empirical values
of the degrees and those estimated by the different kinds of distributions.}

\end{figure}

\subsection*{Waste type II}

\begin{table}[H]
\begin{centering}
\begin{tabular}{|c|c|c|c|c|c|}
\hline 
No. & distribution & NlogL & BIC & AIC & AICc\tabularnewline
\hline 
\hline 
1 & generalized pareto & -3088.32 & -6161.10 & -6170.63 & -6170.49\tabularnewline
\hline 
2 & generalized extreme value & -1764.62 & -3513.72 & -3523.25 & -3523.11\tabularnewline
\hline 
3 & t-location scale & -1386.06 & -2756.60 & -2766.13 & -2765.99\tabularnewline
\hline 
4 & exponential & -739.19 & -1473.18 & -1476.36 & -1476.33\tabularnewline
\hline 
5 & logistic & -535.85 & -1061.34 & -1067.69 & -1067.63\tabularnewline
\hline 
6 & normal & -422.29 & -834.22 & -840.57 & -840.50\tabularnewline
\hline 
7 & extreme value & -287.85 & -565.35 & -571.71 & -571.64\tabularnewline
\hline 
\end{tabular}
\par\end{centering}
\caption{Values of the statistical parameters quantifying the goodness of fit
of the distributions fitting the in-strength of the nodes of the W4
network of wastes type II. Only the top seven distributions are shown.}
\end{table}

\begin{table}[H]
\begin{centering}
\begin{tabular}{|c|c|c|c|c|c|}
\hline 
No. & distribution & NlogL & BIC & AIC & AICc\tabularnewline
\hline 
\hline 
1 & generalized extreme value & -1051.89 & -2088.25 & -2097.78 & -2097.64\tabularnewline
\hline 
2 & generalized pareto & -1048.39 & -2081.26 & -2090.78 & -2090.64\tabularnewline
\hline 
3 & t-location scale & -925.87 & -1836.21 & -1845.74 & -1845.60\tabularnewline
\hline 
4 & beta & -896.47 & -1782.60 & -1788.95 & -1788.88\tabularnewline
\hline 
5 & exponential & -739.18 & -1473.18 & -1476.36 & -1476.33\tabularnewline
\hline 
6 & logistic & -555.88 & -1101.41 & -1107.76 & -1107.69\tabularnewline
\hline 
7 & normal & -444.52 & -878.68 & -885.04 & -884.97\tabularnewline
\hline 
\end{tabular}
\par\end{centering}
\caption{Values of the statistical parameters quantifying the goodness of fit
of the distributions fitting the out-strength of the nodes of the
W4 network of wastes type II. Only the top seven distributions are
shown.}
\end{table}

\begin{figure}
\begin{centering}
\includegraphics[width=1\textwidth]{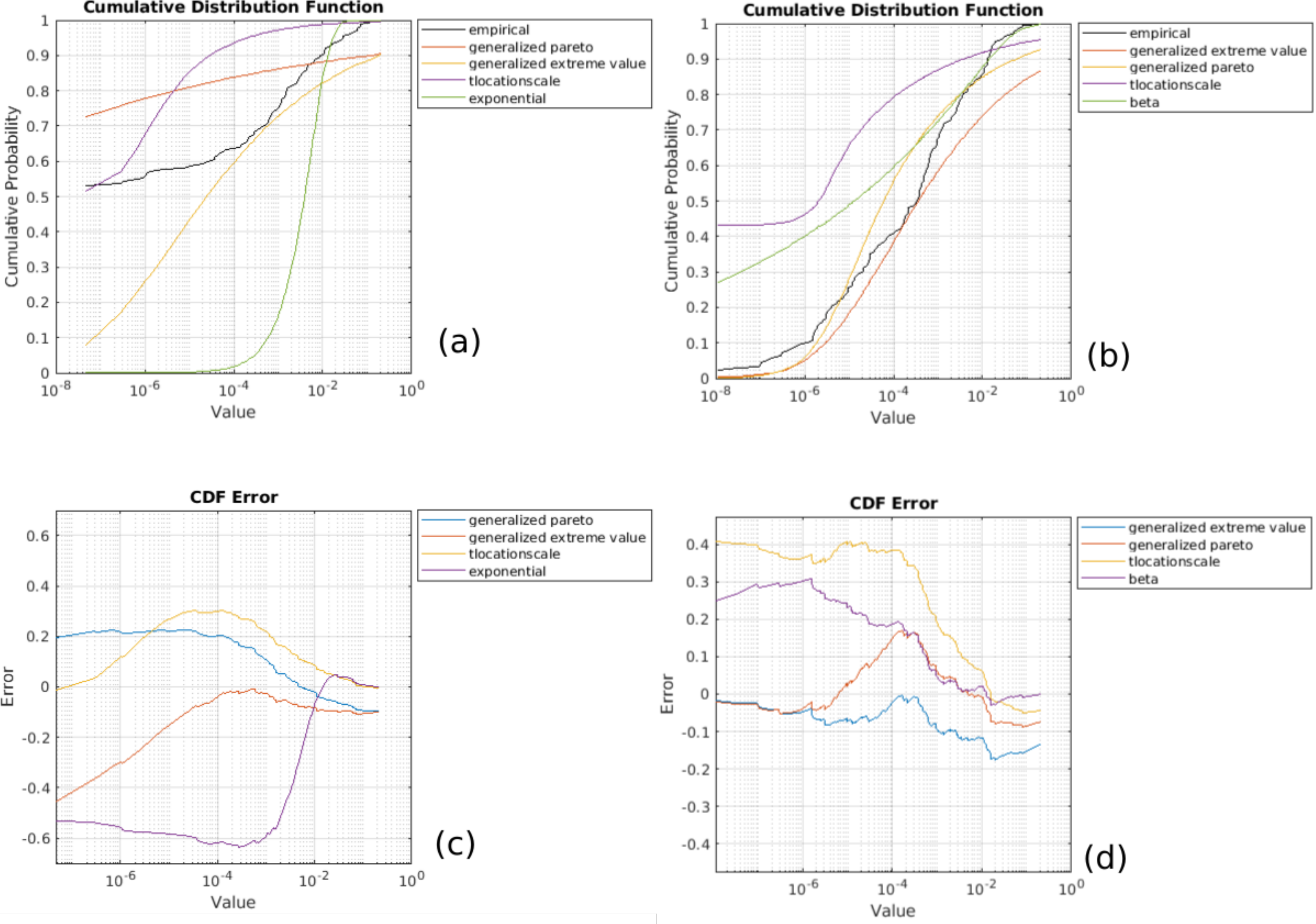}
\par\end{centering}
\caption{Cumulative in- (a) and out-degree (b) distributions and the corresponding errors (c)-(d) for the W4 of type II
wastes. The empirical distribution refers to the data of the W4 network
and the others correspond to the best fits using different kinds of
distributions, e.g., lognormal, loglogistic, generalized extreme value,
etc. The errors are obtained as the differences of the empirical values
of the degrees and those estimated by the different kinds of distributions.}
\end{figure}

\subsection*{Waste type III}

\begin{table}[h]
\begin{centering}
\begin{tabular}{|c|c|c|c|c|c|}
\hline 
No. & distribution & NlogL & BIC & AIC & AICc\tabularnewline
\hline 
\hline 
1 & generalized pareto & -405.46 & -799.25 & -804.93 & -804.40\tabularnewline
\hline 
2 & generalized extreme value & -279.99 & -548.31 & -553.99 & -553.45\tabularnewline
\hline 
3 & t-location scale & -241.62 & -471.56 & -477.23 & -476.70\tabularnewline
\hline 
4 & exponential & -141.70 & -279.51 & -281.40 & -281.31\tabularnewline
\hline 
5 & logistic & -88.38 & -168.98 & -172.77 & -172.51\tabularnewline
\hline 
6 & normal & -66.97 & -126.15 & -129.94 & -129.68\tabularnewline
\hline 
7 & beta & -54.64 & -101.50 & -105.29 & -105.02\tabularnewline
\hline 
\end{tabular}
\par\end{centering}
\caption{Values of the statistical parameters quantifying the goodness of fit
of the distributions fitting the in-strength of the nodes of the W4
network of wastes type III. Only the top seven distributions are shown.}
\end{table}

\begin{table}[h]
\begin{centering}
\begin{tabular}{|c|c|c|c|c|c|}
\hline 
No. & distribution & NlogL & BIC & AIC & AICc\tabularnewline
\hline 
\hline 
1 & generalized pareto & -292.38 & -573.10 & -578.77 & -578.24\tabularnewline
\hline 
2 & generalized extreme value & -264.19 & -516.70 & -522.38 & -521.85\tabularnewline
\hline 
3 & t-location scale & -222.87 & -434.06 & -439.73 & -439.20\tabularnewline
\hline 
4 & exponential & -141.70 & -279.51 & -281.40 & -281.31\tabularnewline
\hline 
5 & beta & -116.92 & -226.06 & -229.84 & -229.58\tabularnewline
\hline 
6 & logistic & -92.51 & -177.24 & -181.03 & -180.77\tabularnewline
\hline 
7 & normal & -72.19 & -136.60 & -140.38 & -140.12\tabularnewline
\hline 
\end{tabular}
\par\end{centering}
\caption{Values of the statistical parameters quantifying the goodness of fit
of the distributions fitting the out-strength of the nodes of the
W4 network of wastes type III. Only the top seven distributions are
shown.}
\end{table}

\begin{figure}
\begin{centering}
\includegraphics[width=1\textwidth]{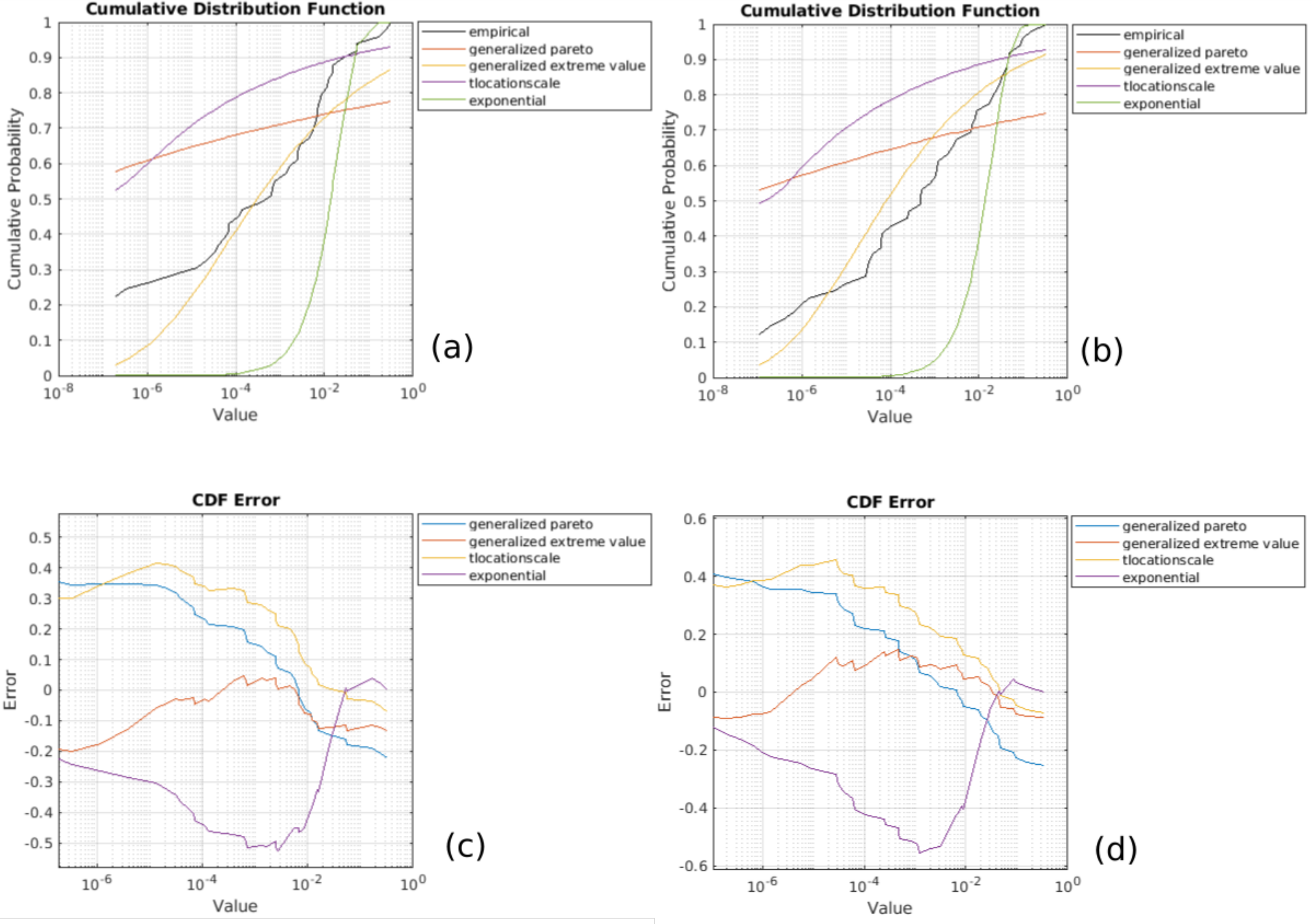}
\par\end{centering}
\caption{Cumulative in- (a) and out-degree (b) distributions and the corresponding errors (c)-(d) for the W4 of type III
wastes. The empirical distribution refers to the data of the W4 network
and the others correspond to the best fits using different kinds of
distributions, e.g., lognormal, loglogistic, generalized extreme value,
etc. The errors are obtained as the differences of the empirical values
of the degrees and those estimated by the different kinds of distributions.}
\end{figure}

\section*{Chemical fingerprints common in hazardous waste}

It has been argued that all chemical substances which are outside
their usual environments or at concentrations above normal represent
a contaminant and that they become pollutants when accumulations are
sufficient to affect the environment or living organisms \cite{General_1}.
The sources of these chemicals may be very diverse, but some of the
top contaminants emerge from waste. These are the cases of heavy
metals, such as arsenic, lead, mercury, hexavalent chromium, and cadmium; volatile organic compounds like vinylchloride, benzene, hexachlorobutadiene
or persistent organic pollutants like polycyclic aromatic compounds,
polychlorinated biphenyls and dioxins \cite{General_1}. The distribution
of contaminants across the world is very unequal with a bigger impact
on low and middle-income countries \cite{General_2}. In these countries
there are serious problems with the disposal of waste, which is poorly
managed, regulated or controlled \cite{General_2}, and obsolete techniques
are applied for their processing, which together with the lack of
governmental infrastructure makes the situation critical \cite{General_2}.
The substantial inform ``Chemical Pollution in Low and Middle-Income
Countries'' \cite{General_2} has identified many of the problems
emerging in these countries due to the chemical pollution in spite
of the many internationally existing legislation, such as the Stockholm,
Basel, and Rotterdam Conventions, and the Strategic Approach to International
Chemicals Management (SAICM).

Generally, the term \textquotedblleft hazardous\textquotedblright{}
waste (HW) is used to define waste with potential threats to public
health \cite{General_4} and the environment. It includes waste electrical
and electronic equipment, commonly designated as e-waste, which has
become the fastest-growing component of the solid-waste streams in
the world \cite{General_3}. E-waste disposal and its informal management
generates highly toxic heavy metals, brominated flame retardants,
non-dioxin-like polychlorinated biphenyls (PCB), polycyclic aromatic
hydrocarbons (PAH), polychlorinated dibenzo-p-dioxins (PCDD), polychlorinated
dibenzofurans (PBDF) and dioxin-like polychlorinated biphenyls (DL-PCB).
These compounds are endocrine disrupters, and most are neuro- and
immune-toxic as well. Another important source of chemical contamination
is the clinical and medical waste \cite{General_5,General_6,General_7}.
Some attention has been given to the cases of infection transmission
due to the inappropriate disposal and handling of this type of waste
\cite{General_5,General_6,General_7}, but less is known about the
chemical traces left by medical waste on the environment and human
populations. Another growing source of chemical contaminants is municipal
solid waste, which tripled from 1965 to 2015 \cite{General_8}.

Waste problems in developing countries is aggravated by the transboundary
trade of HW, which increases their burden of certain kinds of waste
in those countries \cite{General_10,General_11,General_12,General_13,General_14}.
The problem should be analyzed from a wide perspective. Namely, we
do not claim here that waste trade is the source of all the environmental
and human health problems that waste produces. We claim that in those
countries with a large burden due to bad practices and poor resources
for waste management, importing any amount of waste will only aggravate
their situation. This is illustrated by the fact that in Africa there
are significant levels of soil pollution due to agricultural activities,
mining, roadside emissions, auto-mechanic workshops, their own refuse
dumps and e-waste \cite{Africa_1}. Then, when poor African countries
like Nigeria, become a dumping ground for HW imported from abroad
\cite{Africa_2}, the situation become of a very high risk. In Africa
there are 67,740 health-care facilities which generate 56,100-487,100
tonnes of medical waste per year. These global amounts are not exaggeratedly
large, but the risk is very high considering that medical waste is
rarely sorted which makes the amounts of HW much higher than in other
parts of the world \cite{Africa_3,Africa_4}. Then, when these countries
receive large amounts of e-waste from abroad they do not have resources
for dealing with its recycling. The consequences are elevated levels
of e-waste pollutants in water, air, soil, dust, fish, vegetable,
and human blood, urine, breast milk, producing headache, cough and
chest pain, stomach discomfort, miscarriage, abnormal thyroid and
reproductive function, reduction of gonadal hormone, and cancer in
those involved with the processing of e-waste \cite{Africa_5}.

In Table \ref{fingerprints} we report the chemical fingerprints left
by different types of waste according to an intensive bibliographic
search carried out in this work. In Tables \ref{Waste_type_I}, \ref{Waste_type_II}
and \ref{Waste_type_III} we then report every individual waste category
reported by the Basel Convention, which are on the types I-III considered
in this work. We report the chemical fingerprints left by these waste
based on the specific report on their contents at the Basel Convention
web page.

\begin{table}[H]
\begin{centering}
\begin{tabular}{|c|>{\raggedright}p{8cm}|>{\raggedright}p{1.7cm}|}
\hline 
fingerprint & wastes & ref.\tabularnewline
\hline 
\hline 
HM & pesticides, paints and pigments, enamel, varnishes, dyes, catalysts,
batteries, accumulators, printing products, e-waste, metal products,
asbestos, anticorrosive, technical oils, sewage sludge, waste incineration
products, waste from plastic production, PVC plastics, colored glass,
glues, ash of coal, metalurgical slag, galvanic waste, waste of nonferrous
metallurgy, waste of leather industry, agriculture waste, waste of
medicines, medical/clinical waste & \cite{HM_1,HM_2,E-waste_Chemistry,Clinical solid waste,HM_3,HM_4,Mineral_Oils,HM_5}\tabularnewline
\hline 
VOC & solvents, wastes from petroleum refining, synthetic resin, textile
dyeing and printing, leather manufacturing, the pharmaceutical industry,
pesticide manufacturing, coating, printing ink, adhesive manufacturing,
spraying, printing, e-waste, plastic solid waste recycling, municipal
solid waste, agricultural waste burning & \cite{E-waste_Chemistry,Clinical solid waste,VOC_1,VOC_2,VOC_3,VOC_4,VOC_5,VOC_6,VOC_7}\tabularnewline
\hline 
POP & municipal solid waste, medical waste, sewage sludge, and hazardous
waste incinerations; informal recycling of e-waste; drilling wastes;
PCB-containing additives in rubber, resins, carbonless copy paper,
inks, hydraulic fluids, heat-transfer fluids, plasticizers, and lubricants;
PCB-containing transformers, capacitors; PCB-containing wastes from
coating of papers, sealants of cars, coloring of China glassware,
color television parts, the effect extension agents of agricultural
chemicals, oil additive agents & \cite{E-waste_Chemistry,Clinical solid waste,Mineral_Oils,POP_1,POP_2,POP_3,POP_4,POP_5}\tabularnewline
\hline 
\end{tabular}
\par\end{centering}
\caption{Chemical fingerprints left by waste in the environment and/or human
health. The fingerprints are classified as heavy metals (HM), volatile
organic compounds (VOC) and persistent organic pollutants (POP).}

\label{fingerprints}
\end{table}

\subsection*{Waste categories in types I-III}

\begin{table}[H]
\begin{centering}
\resizebox{12.1cm}{!} {
\begin{tabular}{|c|>{\raggedright}p{8cm}|>{\raggedright}p{1.7cm}|}
\hline 
category & description & fingerprints\tabularnewline
\hline 
\hline 
Y1 & Clinical wastes from medical care in hospitals, medical centers and
clinics & HM, VOC, POP\tabularnewline
\hline 
Y2 & Wastes from the production and preparation of pharmaceutical products & HM, VOC, POP\tabularnewline
\hline 
Y3 & Waste pharmaceuticals, drugs and medicines & HM, VOC, POP\tabularnewline
\hline 
Y4 & Wastes from the production, formulation and use of biocides and phytopharmaceuticals & HM, VOC, POP\tabularnewline
\hline 
Y5 & Wastes from the manufacture, formulation and use of wood preserving
chemicals & HM, VOC, POP\tabularnewline
\hline 
Y6 & Wastes from the production, formulation and use of organic solvents & VOC\tabularnewline
\hline 
Y7 & Wastes from heat treatment and tempering operations containing cyanides & VOC\tabularnewline
\hline 
Y8 & Waste mineral oils unfit for their originally intended use & VOC, POP\tabularnewline
\hline 
Y9 & Waste oils/water, hydrocarbons/water mixtures, emulsions & VOC, POP\tabularnewline
\hline 
Y10 & Waste substances and articles containing or contaminated with polychlorinated
biphenyls (PCBs) and/or polychlorinated terphenyls (PCTs) and/or polybrominated
biphenyls (PBBs) & POP\tabularnewline
\hline 
Y11 & Waste tarry residues arising from refining, distillation and any pyrolytic
treatment & HM, VOC, POP\tabularnewline
\hline 
Y12 & Wastes from production, formulation and use of inks, dyes, pigments,
paints, lacquers, varnish & VOC, POP\tabularnewline
\hline 
Y13 & Wastes from production, formulation and use of resins, latex, plasticizers,
glues/adhesives & VOC, POP\tabularnewline
\hline 
Y14 & Waste chemical substances arising from research and development or
teaching activities which are not identified and/or are new and whose
effects on man and/or the environment are not known & HM, VOC, POP\tabularnewline
\hline 
Y15 & Wastes of an explosive nature not subject to other legislation & VOC\tabularnewline
\hline 
Y16 & Wastes from production, formulation and use of photographic chemicals
and processing materials & HM, VOC, POP\tabularnewline
\hline 
Y17 & Wastes resulting from surface treatment of metals and plastics & VOC, POP\tabularnewline
\hline 
Y18 & Residues arising from industrial waste disposal operations & HM, VOC, POP\tabularnewline
\hline 
\end{tabular}
}
\par\end{centering}
\caption{Waste categories included in the Basel Convention which are grouped
in the type I of wastes and the chemical fingerprints (CF) left by
them in the environment and/or human health: heavy metals (HM), volatile
organic compounds (VOC) and persistent organic pollutants (POP).}

\label{Waste_type_I}
\end{table}

\begin{table}[H]
\begin{centering}
\begin{tabular}{|c|>{\raggedright}p{8cm}|>{\raggedright}p{1.7cm}|}
\hline 
category & description & fingerprints\tabularnewline
\hline 
\hline 
Y19 & Metal carbonyls & HM, VOC\tabularnewline
\hline 
Y20 & Beryllium; beryllium compounds & HM\tabularnewline
\hline 
Y21 & Hexavalent chromium compounds & HM\tabularnewline
\hline 
Y22 & Copper compounds & HM, VOC, POP\tabularnewline
\hline 
Y23 & Zinc compounds & HM\tabularnewline
\hline 
Y24 & Arsenic; arsenic compounds & HM\tabularnewline
\hline 
Y25 & Selenium; selenium compounds & HM\tabularnewline
\hline 
Y26 & Cadmium; cadmium compounds & HM\tabularnewline
\hline 
Y27 & Antimony; antimony compounds & HM\tabularnewline
\hline 
Y28 & Tellurium; tellurium compounds & HM\tabularnewline
\hline 
Y29 & Mercury; mercury compounds & HM\tabularnewline
\hline 
Y30 & Thallium; thallium compounds & HM\tabularnewline
\hline 
Y31 & Lead; lead compounds & HM\tabularnewline
\hline 
Y32 & Inorganic fluorine compounds excluding calcium fluoride & HM\tabularnewline
\hline 
Y33 & Inorganic cyanides & VOC\tabularnewline
\hline 
Y34 & Acidic solutions or acids in solid form & VOC, POP\tabularnewline
\hline 
Y35 & Basic solutions or bases in solid form & VOC, POP\tabularnewline
\hline 
Y36 & Asbestos (dust and fibres) & \tabularnewline
\hline 
Y37 & Organic phosphorus compounds & VOC, POP\tabularnewline
\hline 
Y38 & Organic cyanides & VOC\tabularnewline
\hline 
Y39 & Phenols; phenol compounds including chlorophenols & VOC\tabularnewline
\hline 
Y40 & Ethers & VOC\tabularnewline
\hline 
Y41 & Halogenated organic solvents & VOC, POP\tabularnewline
\hline 
Y42 & Organic solvents excluding halogenated solvents & VOC\tabularnewline
\hline 
Y43 & Any congenor of polychlorinated dibenzo-furan & POP\tabularnewline
\hline 
Y44 & Any congenor of polychlorinated dibenzo-p-dioxin & POP\tabularnewline
\hline 
Y45 & Organohalogen compounds other than substances referred to in this
Annex (e.g. Y39, Y41, Y42, Y43, Y44) & HM, VOC, POP\tabularnewline
\hline 
\end{tabular}
\par\end{centering}
\caption{Waste categories included in the Basel Convention which are grouped
in the type II of wastes and the chemical fingerprints (CF) left by
them in the environment and/or human health: heavy metals (HM), volatile
organic compounds (VOC) and persistent organic pollutants (POP).}

\label{Waste_type_II}
\end{table}

\begin{table}[H]
\begin{centering}
\begin{tabular}{|c|>{\raggedright}p{8cm}|>{\raggedright}p{1.7cm}|}
\hline 
category & description & fingerprints\tabularnewline
\hline 
\hline 
Y46 & Wastes collected from households & HM, VOC, POP\tabularnewline
\hline 
Y47 & Residues arising from the incineration of household wastes & HM, VOC, POP\tabularnewline
\hline 
\end{tabular}
\par\end{centering}
\caption{Waste categories included in the Basel Convention which are grouped
in the type III of wastes and the chemical fingerprints (CF) left
by them in the environment and/or human health: heavy metals (HM),
volatile organic compounds (VOC) and persistent organic pollutants
(POP).}

\label{Waste_type_III}
\end{table}

\section*{Centrality of countries in the W4 networks}
\begin{figure}[H]
\begin{centering}
\includegraphics[width=11cm]{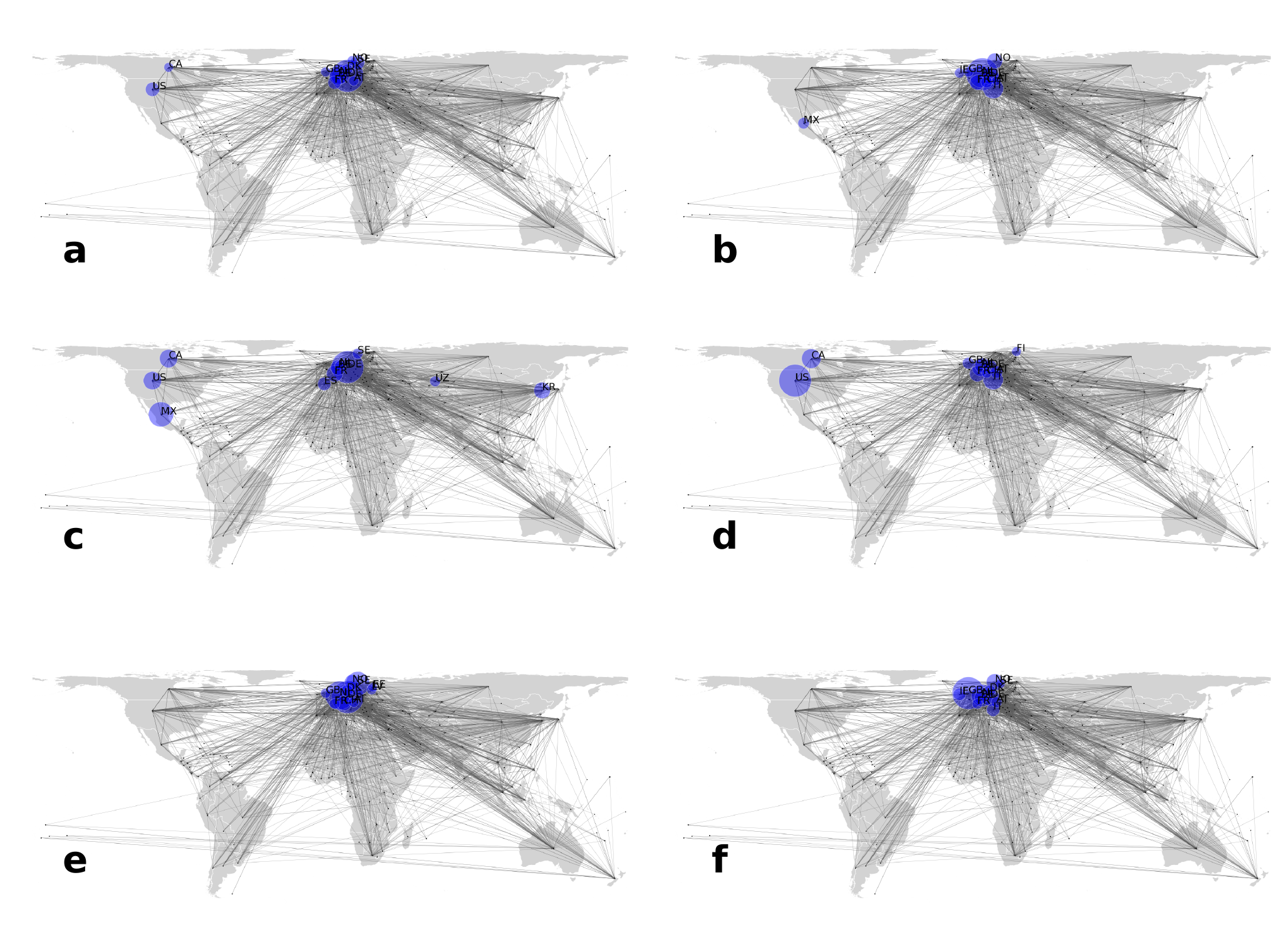}
\par\end{centering}
\caption{Main importers (left panels) and exporters (right panels) for wastes
of types I (a, b), II (c, d) and III (e, f).These values correspond
to the in- and out-strengths (weighted degrees) of the corresponding
countries in the respective W4 networks. Here, we highlight the first twelve importers and exporters. Country codes belong to the ISO-alpha-2 standard. Map tiles by Bjorn Sandvik, under CC BY-SA 3.0 available at http://thematicmapping.org/downloads/world\_borders.php.} \label{strengths maps}
\end{figure}

\begin{figure}
\begin{centering}
\includegraphics[width=11cm]{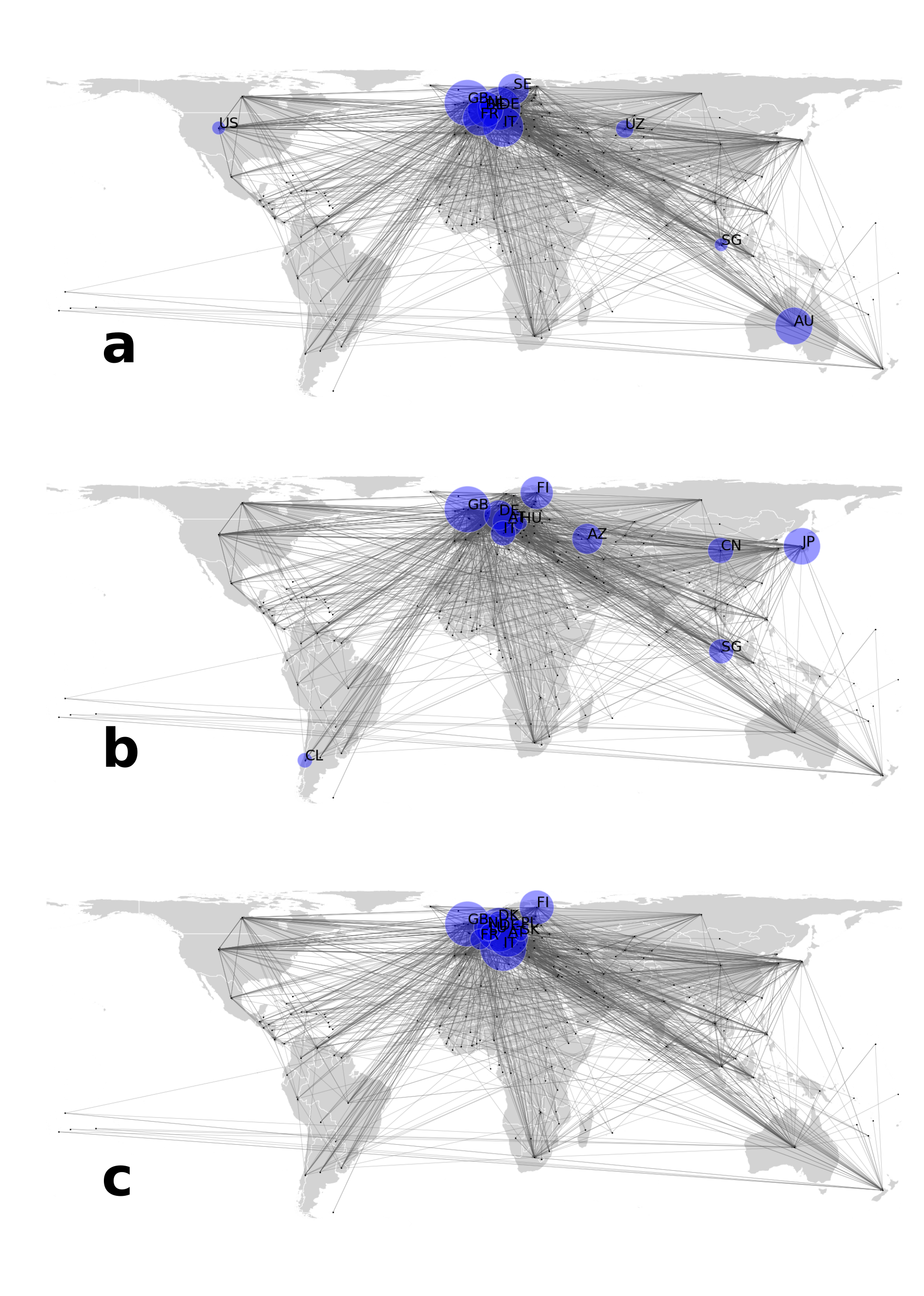}
\par\end{centering}
\caption{Main countries to which more flux of wastes pass through them
for types I (a), II (b) and III (c). These values correspond
to the betweenness centrality of the corresponding
countries in the respective W4 networks. Map tiles by Bjorn Sandvik, under CC BY-SA 3.0 available at http://thematicmapping.org/downloads/world\_borders.php.} \label{betweenness maps}
\end{figure}

\begin{figure}
\begin{centering}
\includegraphics[width=1\textwidth]{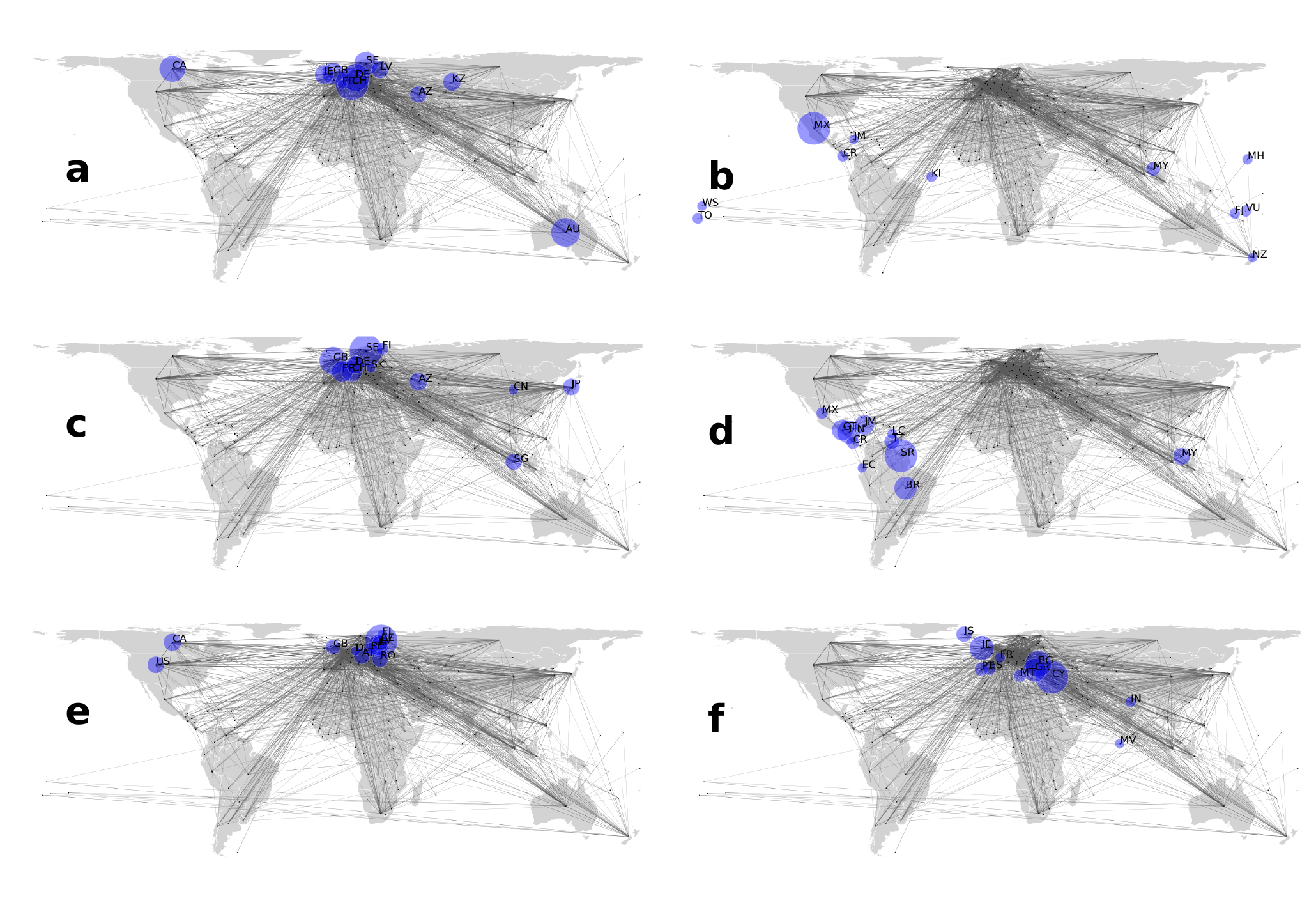}
\par\end{centering}
\caption{Main countries to which the traffic of the network is closer from all other territories in the network. Values correspond to In-closeness and Out-closeness centrality for type I (a,b), II (c,d), II (e,f). Main countries which are closer to all countries/territories in the waste network for out-closeness centrality of type III (c). We highlight the first twelve hubs of each networks. Country codes belong to the ISO-alpha-2 standard. Map tiles by Bjorn Sandvik, under CC BY-SA 3.0 available at http://thematicmapping.org/downloads/world\_borders.php.} \label{closeness maps}
\end{figure}

Here we introduce other networks metrics for waste types I, II, and III. Fig. \ref{strengths maps} depicts the in-/out-strengths. Fig. \ref{betweenness maps} shows the betweenness centrality, and Fig. \ref{closeness maps} introduces the in-closeness for types I, II, III; and the out-closeness.
\begin{figure}[H]
\begin{centering}
\includegraphics[width=1\textwidth]{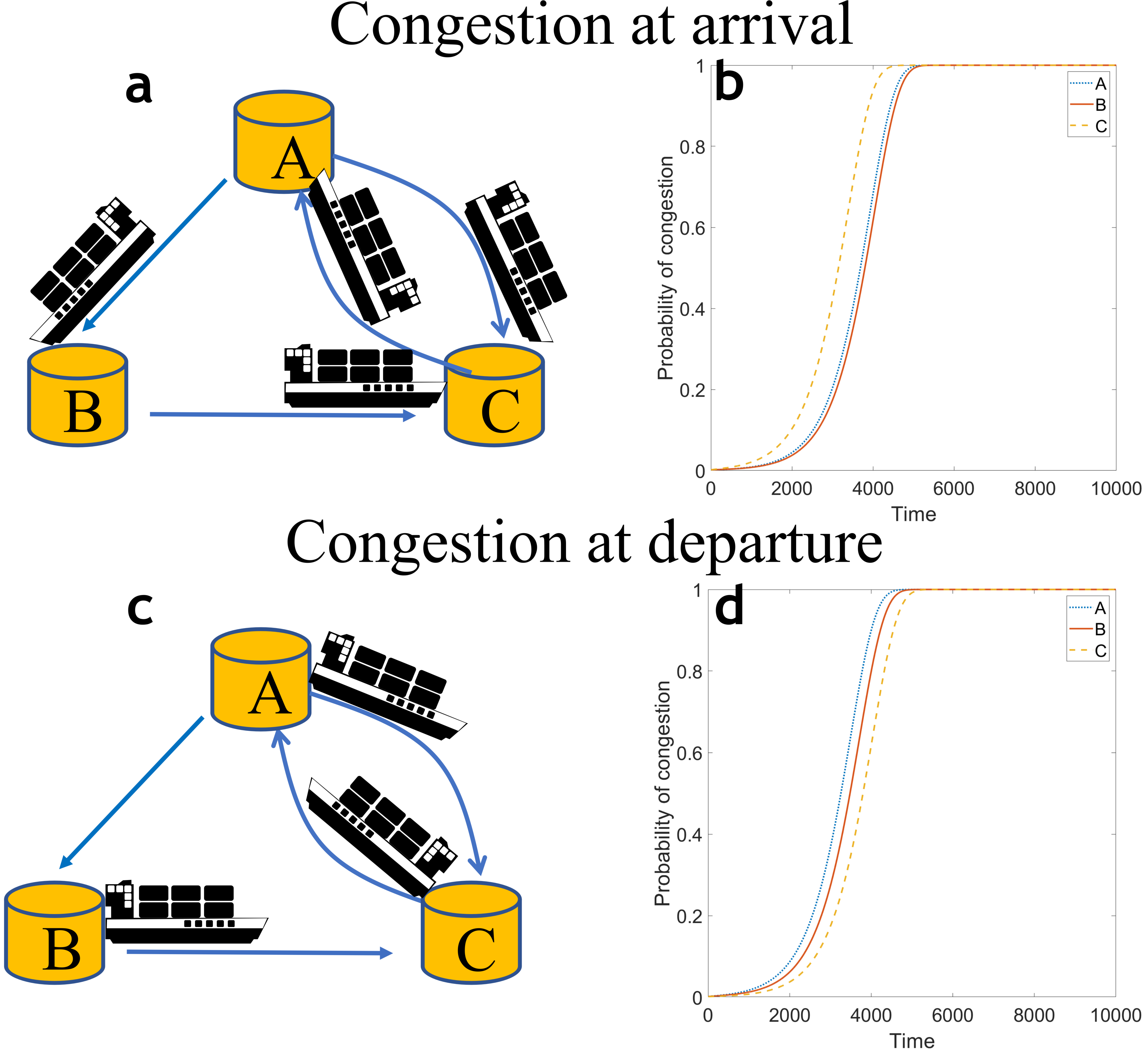}
\par\end{centering}
\caption{Schematic illustration of the ``congestion at arrival" (\textbf{a}) and
``congestion at departure" (\textbf{c}) models and the time-evolution of the
congestion propagation through the nodes using these models (\textbf{b} and \textbf{d}).
Notice that in the congestion at arrival (panel \textbf{b}), node C reaches 50 \%
of congestion at a earlier time than A and B. In the congestion at
departure (panel \textbf{d}), node A reaches 50 \% of congestion earlier than B and
C. Also notice that the ordering of congestion times at departure and
arrival are not simply one the reverse of the other.}
\label{congestion_models}
\end{figure}

\section*{Theoretical modeling approach}
The logistic dynamic model on a network is written as (see \cite{Bullo}
for analysis in the case of Susceptible-Infected model, which is a
particular case of the general model written here):

\begin{equation}
\dfrac{dw_{i}\left(t\right)}{dt}=\beta\left(1-w_{i}\left(t\right)\right)\sum_{j=1}^{n}A_{ij}w_{j}\left(t\right),t\geq t_{0},\label{eq:SI_original}
\end{equation}
where $A_{ij}$ are the entries of the adjacency matrix of the W4
for the pair of countries $i$ and $j$. In matrix-vector form it
becomes:

\begin{equation}
\dfrac{d\mathbf{w}\left(t\right)}{dt}=\beta\left[I_{N}-\textnormal{diag}\left(\mathbf{w}\left(t\right)\right)\right]A\mathbf{w}\left(t\right),
\end{equation}
 with initial condition $\mathbf{w}\left(0\right)=\mathbf{w}_{0},$where $I_{N}$ is
the identity matrix of order $N$. This model can be rewritten as

\begin{equation}
\dfrac{1}{1-w_{i}(t)}\dfrac{dw_{i}\left(t\right)}{dt}=\beta\sum_{j=1}^{n}A_{ij}\left(1-e^{-\left(-\log\left(1-w_{j}\left(t\right)\right)\right)}\right),
\end{equation}
which is equivalent to

\begin{equation}
\dfrac{dy_{i}\left(t\right)}{dt}=\beta\sum_{j=1}^{n}A_{ij}f\left(y_{j}\left(t\right)\right),
\end{equation}
where $y_{i}\left(t\right)\coloneqq g\left(x_{i}\left(t\right)\right)=-\log\left(1-w_{i}\left(t\right)\right)\in\left[0,\infty\right]$,
$f\left(y\right)\coloneqq1-e^{-y}=g^{-1}\left(y\right)$.

We now consider a dynamics with non-locality by time or dynamic memory
as described in the main text. To write the logistic model in this
new context we start as follows. Let $0<\alpha<1$, then

\[
\left\{ \begin{array}{l}
{\displaystyle \int_{0}^{t}g_{1-\alpha}\left(t-\tau\right)\dfrac{\left(1-w_{i}\right)'\left(\tau\right)}{w_{i}\left(\tau\right)}\,d\tau=-\beta^{\alpha}\left(1-w_{i}\right)\left(t\right),}\\
\\
{\displaystyle \int_{0}^{t}g_{1-\alpha}\left(t-\tau\right)\dfrac{w_{i}'\left(\tau\right)}{\left(1-w_{i}\right)\left(\tau\right)}\,d\tau=\beta^{\alpha}w_{i}\left(t\right).}
\end{array}\right.
\]
Then, we have 
\begin{equation}
\int_{0}^{t}g_{1-\alpha}\left(t-\tau\right)\dfrac{w_{i}'\left(\tau\right)}{1-w_{i}\left(\tau\right)}\,d\tau=\beta^{\alpha}w_{i}\left(t\right),
\end{equation}
which for the case of a network is written as
\begin{equation}
\begin{array}{ll}
{\displaystyle \int_{0}^{t}g_{1-\alpha}\left(t-\tau\right)\dfrac{w_{i}'\left(\tau\right)}{1-w_{i}\left(\tau\right)}\,d\tau=\beta^{\alpha}\sum_{j=1}^{n}A_{ij}w_{j},} & t>0,w_{i}(0)\in[0,1].\end{array}\label{fractsystem}
\end{equation}
We can rewrite (\ref{fractsystem}) in a matrix-vector form: 
\begin{equation}
D_{t}^{\alpha}(-\log(\mathbf{1}-\mathbf{w}))(t)=\beta^{\alpha}A\mathbf{w}\left(t\right),
\end{equation}
with the logarithm taken entrywise, and with initial condition $\mathbf{w}\left(0\right)=\mathbf{w}_{0}.$

In order to solve analytically the previous equation we apply the
Lee-Tenneti-Eun (LTE) transformation \cite{Lee_et_al} which produces
the following linearized equation 
\begin{equation}
D_{t}^{\alpha}\hat{\mathbf{y}}\left(t\right)=\beta^{\alpha}A\text{diag}\left(\mathbf{1}-\mathbf{w}_{0}\right)\hat{\mathbf{y}}\left(t\right)+\beta^{\alpha}A\mathbf{b}\left(\mathbf{w}_{0}\right),\label{eqlineal}
\end{equation}
where $\hat{\mathbf{w}}\left(t\right)=f\left(\hat{\mathbf{y}}\left(t\right)\right)$
in which $\hat{\mathbf{w}}\left(t\right)$ is an approximate solution to the
fractional SI model, $\hat{\mathbf{y}}$ is the solution of (\ref{eqlineal})
with initial condition $\hat{\mathbf{y}}\left(0\right)=g\left(\mathbf{x}\left(0\right)\right)$, $\mathbf{1}$ is the all-ones vector, 
and $\mathbf{b}\left(\mathbf{\mathbf{w}}\right):=\mathbf{w}+\left(\mathbf{1}-\mathbf{w}\right)\log\left(\mathbf{1}-\mathbf{w}\right).$
For convenience, we write $\Omega:=\text{diag}\left(\mathbf{1}-\mathbf{w}_{0}\right),$
and $\hat{A}=A\Omega.$ Then, we have proved that this approximate
solution $\hat{\mathbf{w}}\left(t\right)$ is a non-divergent upper bound to
the exact solution $\mathbf{x}(t)$.

\begin{thm}
For any $t\geq0$, we have 
\[
\mathbf{w}(t)\preceq\hat{\mathbf{w}}(t)=f(\hat{\mathbf{y}}(t)),
\]
under the same initial conditions $w_{0}:=w(0)=\hat{w}(0),$ where
the solution $\hat{\mathbf{y}}$ of (\ref{eqlineal}) is given by 
\begin{equation}
\hat{\mathbf{y}}\left(t\right)=E_{\alpha,1}\left((\beta t)^{\alpha}\hat{A}\right)g\left(\mathbf{w}_{0}\right)+\sum_{n=0}^{\infty}\frac{(\beta t)^{\alpha\left(n+1\right)}\hat{A}^{n}A\mathbf{b}\left(\mathbf{w}_{0}\right)}{\Gamma\left(\alpha\left(n+1\right)+1\right)}.\label{y}
\end{equation}
Furthermore, $\|\hat{\mathbf{w}}(t)-\mathbf{w}(t)\|\to0$ and $\|\tilde{\mathbf{w}}(t)-\mathbf{w}(t)\|\to\infty$
as $t$ goes to infinity.
\end{thm}

We also proved that when all values of the initial condition are smaller
than one, i.e., $\mathbf{w}_{0}\preceq\mathbf{1}$, which means that at the starting
point of the simulation no country is completely congested of waste,
the solution of the fractional logistic waste congestion model is
\begin{equation}
\hat{\mathbf{y}}\left(t\right)=g\left(\mathbf{w}_{0}\right)+\left[E_{\alpha,1}\left((\beta t)^{\alpha}\hat{A}\right)-I\right]\Omega^{-1}\mathbf{w}\left(0\right).
\end{equation}

This is important because if we consider the plausible case that the
probability of getting congested at $t=0$ is the same for every country,
which mathematically is written as: $w_{0}=\frac{c}{N}$ where $c\in\mathbb{R}^{+},$
we have that

\begin{align}
\hat{\mathbf{y}}\left(t\right) & =\left(\frac{1-\gamma}{\gamma}\right)E_{\alpha,1}\Bigl(t^{\alpha}\beta^{\alpha}\gamma A\Bigr)\vec{\mathbf{1}}-\left(\frac{1-\gamma}{\gamma}+\log\gamma\right){\mathbf{1}},
\end{align}
where $\gamma=1-w_{0}$ and we have used the fact that $\textnormal{diag}\left(\mathbf{1}-\mathbf{w}\left(0\right)\right)=\gamma I$,
where $I$ is the identity matrix.

The Mittag-Leffler function $E_{\alpha,1}\Bigl(\zeta A\Bigr)$ with
$\zeta=\left(\beta t\right)^{1/2}\gamma$, which appears in the approximate
solution of the congestion models described in the main text, belongs
to the class of matrix functions of the adjacency matrix \cite{Matrix_functions}.
It can be written as \cite{ML-1,ML-2,ML-3,ML-4}

\begin{equation}
E_{\alpha,1}\Bigl(\zeta A\Bigr)=\sum_{k=0}^{\infty}\dfrac{\left(\zeta A\right)^{k}}{\Gamma\left(\alpha k+1\right)},\alpha>0.
\end{equation}

If we expand the first terms of this matrix function for a pair of
countries $v$ and $w$ we get:

\begin{equation}
\left(E_{\alpha,1}\Bigl(\zeta A\Bigr)\right)_{vw}=\dfrac{\zeta\left(A\right)_{vw}}{\Gamma\left(\alpha+1\right)}+\dfrac{\zeta^{2}\left(A^{2}\right)_{vw}}{\Gamma\left(2\alpha+1\right)}+\dfrac{\zeta^{3}\left(A^{3}\right)_{vw}}{\Gamma\left(3\alpha+1\right)}+\cdots.
\end{equation}

The first term is different from zero only if the country $v$ exports
some amount of waste to country $w$. The second term accounts for
the export of $v$ to any country $i,$ which then exports to $w$:
$v\rightarrow i\rightarrow w$. The third term accounts for a chain
of the type: $v\rightarrow i\rightarrow j\rightarrow w$ or of an
interchange: $v\rightarrow w\rightarrow v\rightarrow w$. In every
case the amounts exported from one country to another are taken into
account. Notice that such chains could be of infinite length, but
their importance is diminished by the denominator of each particular
term, given by the Euler gamma functions.

A centrality index, like the strengths (in- and out-) only take into
account the contribution of exports/import between pairs of connected
nodes in the network. In case that the country $v$ exports some amount
of waste to country $w$, the out-strength of $v$ is given by the
first term of the previous series expansion. However, this strength
measures do not take into account the chains of lengths longer than
one, such as $v\rightarrow i\rightarrow w$, or $v\rightarrow i\rightarrow j\rightarrow w$.
More importantly, if two countries $v$ and $w$ are not connected
in the network, the strengths measures fail to account for possible
indirect exports/imports between these two countries through an intermediary,
such as in $v\rightarrow i\rightarrow w$.

\begin{figure}[H]
\begin{centering}
\includegraphics[width=1\textwidth]{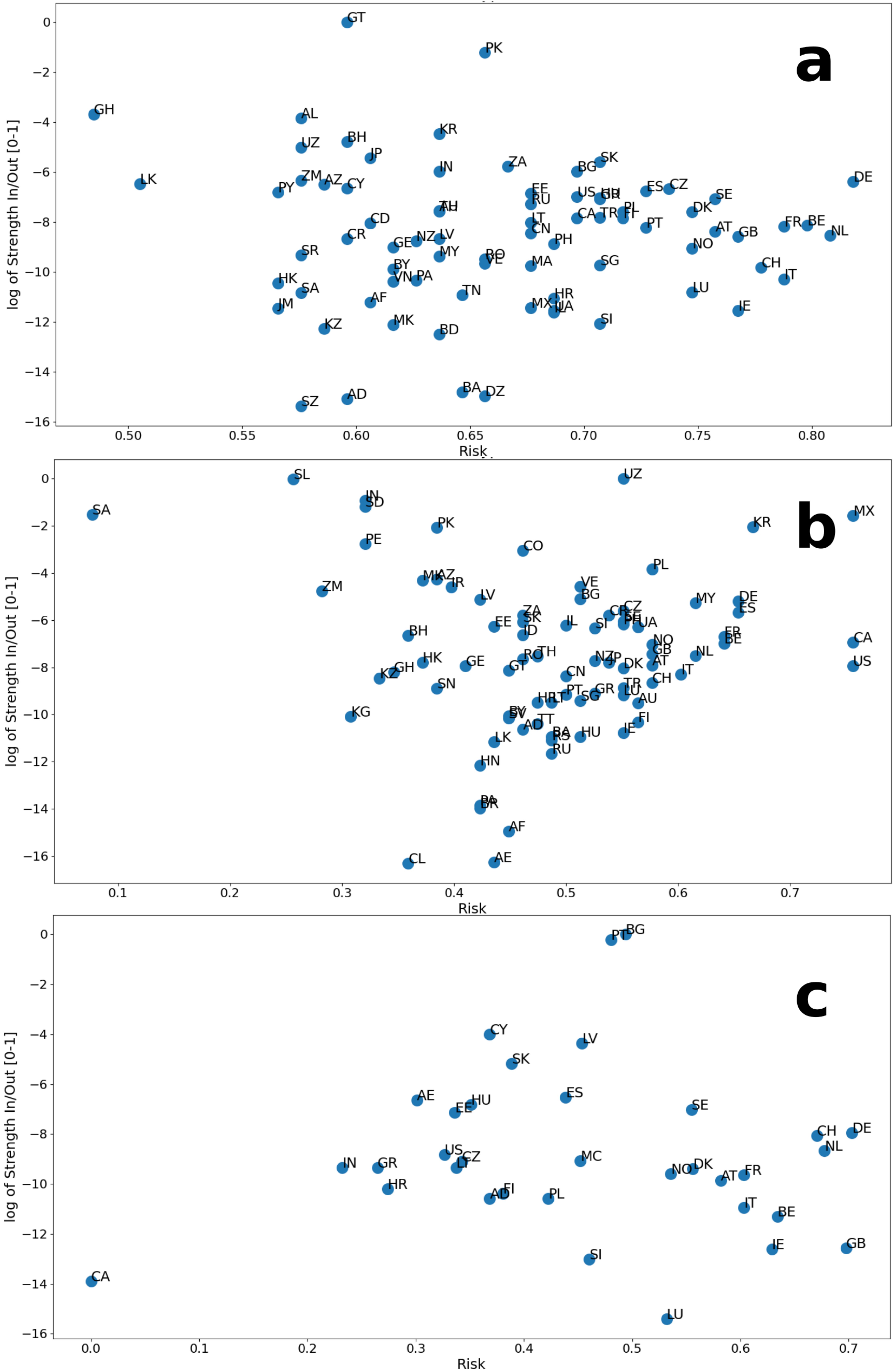}
\par\end{centering}
\caption{Scatter plot between a local measure (strength-in/strength-out), vs congestion risk for countries involved in networks of types I (\textbf{a}) with Pearson index of 0.01, II (\textbf{b}) with Pearson index of 0.00, and III (\textbf{c}).with Pearson index of 0.10 .} \label{spearman}
\end{figure}

This lack of correlation between the first order measures, like strength,
and higher order ones are revealed by the plots (see Fig. \ref{spearman} ) of the ratio of both
strength measures and the index of risk of waste congestion defined
here.

\section*{Chemical fingerprints}

\begin{longtable}[c]{|>{\centering}p{1.2cm}|>{\raggedright}m{8cm}|>{\centering}p{2cm}|}
\hline 
country & Evidence of CF found  & Ref.\tabularnewline
\hline 
\hline 
AF & Reports from foreign military bases on emissions of VOC, PAH, and
HM; 77\% of the volume of waste generated in Kabul are uncollectedmeaning
1180 $m^{3}$ of waste (830 tons) uncollected. & \cite{Afghanistan_1,Afghanistan_2,Afghanistan_3}\tabularnewline
\hline 
BD & \raggedright{}Heavy metal contamination of fruits, vegetables, fish
and ther foodstuffs with risk for human health; serious mismanagement
of medical waste in the capital with serious risk of human health;
Heavy pollution with POP, particularly polychlorinated byphenyls (PCB). & \cite{Bangladesh_1,Bangladesh_2,Bangladesh_3,Bangladesh_4,Bangladesh_5,Bangladesh_6,Bangladesh_7,Bangladesh_8}\tabularnewline
\hline 
BJ & The Benin Republic has the highest per capita hazardous waste generation
in Africa at 65 kg/person/annum; Large amounts of e-waste reported;
1,698 tons of HW incinerated, none recycled nor landfilled; large
exposure of inhabitants of the capital to BTEX. & \cite{Benin_1,Benin_2}\tabularnewline
\hline 
BF & \raggedright{}Heavy metals contamination of soils from informal settlements,
peri-urban agriculture and unregulated waste dumping; problems with
waste management producing heavy metals contamination and with impact
on human health. & \cite{BurkinaFaso_4,BurkinaFaso_2,BurkinaFaso_3,BurkinaFaso_5}\tabularnewline
\hline 
CN & \raggedright{}Heavy metals, VOC and POP pollution due to e-waste disposal
and mismanagement with serious thread for human health; high contamination
levels of PCB from equipments; emission and speciation of VOC from
anthropogenic sources, ncluding high levels of BTEX; high levels of
pollution from medical wastes and their incineration. & \cite{China_1,China_2,China_3,China_4,China_5,China_6,China_7,China_8,China_9,China_10,China_11,China_12,China_13}\tabularnewline
\hline 
CD & \raggedright{}Heavy metals and POP contamination in river, estuary,
and marine sediments from Atlantic Coast; impact of heavy metals on
human health in children and adult populations; contamination of water
resources and food chain by POP. & \cite{DRCongo_1,DRCongo_2,DRCongo_3,DRCongo_4}\tabularnewline
\hline 
DJ & Reports of a shipment of containers with up to 20 metric tons of toxic
chemicals were found leaking in the port of Djibouti with potential
pollution by Arsenic. & \cite{Djibouti_1}\tabularnewline
\hline 
ET & \raggedright{}Problems with uncontrolled waste disposal and heavy
metals contamination of soils and waters in Addis Ababa; contamination
by VOC in urban environment; high levels of pollution by POP, specially
PCB and dioxins at different lcatins and levels of the trophic chain;
seriuos problems of mismanagement of medical wastes with reported
cases of hepatitis B and C directly related to them. & \cite{Ethiopia_1,Ethiopia_2,Ethiopia_3,Ethiopia_4,Ethiopia_5,Ethiopia_6,Ethiopia_7,Ethiopia_8,Ethiopia_9,Ethiopia_10,Ethiopia_11}\tabularnewline
\hline 
GN & \raggedright{}Heavy metals contamination in the Gulf of Guinea. & \cite{Guinea_1,Guinea_2}\tabularnewline
\hline 
IN & \raggedright{}Heavy metals contamination due to diverse wastes, including
e-waste, with serious thread for human health; high levels of VOC
from waste dumps, including high levels of BTEX; problems with dumping
and informal recycling of medical waste with important human health
problems. & \cite{India_1,India_2,India_3,India_4,India_5,India_6,India_7,India_8,India_9,India_10,India_11,India_12}\tabularnewline
\hline 
LS & Open waste combustion emissions of CO2 are estimated to be more than
each country\textquoteright s total national CO2 emissions as reported
by the United Nations; As and Pb concentration levels found in \textit{Cyprinus
carpio} were higher than the WHO permissible limits recommended for
fish consumption Maqalika Reservoir --Maseru, Lesotho; High levels
of disposed e-waste are reported. & \cite{Lesotho_1,Lesotho_2,Lesotho_3}\tabularnewline
\hline 
LR & Waste management activities are getting worse daily due to shortage
of a comprehensive waste management framework; reports of mismanagement
of healthcare waste. & \cite{Liberia_1,Liberia_2}\tabularnewline
\hline 
MG & High levels of HW landfilled (33,812 tons), none recycled and 12,145
tons incinerated; reports of illegal import of car batteries and extraction
of Pb for export. & \cite{Benin_1,Madagascar_1}\tabularnewline
\hline 
MR & \raggedright{}Contamination by POP like PAHs in Atlantic coast of
Mauritania (Levrier Bay Zone); high levels of PCB contamination in
marine ecosystems front of the coast of Mauritania. & \cite{Mauritania_1,Mauritania_2,Mauritania_3}\tabularnewline
\hline 
MX & High levels of BTEX found due among other causes to waste burning;
exposure of children to mixtures of pollutants including PAH, PCB
and HM (As and Pb) in a site with a hazardous waste landfill; reported
levels of emission factors for polychlorinated and polybrominated
dibenzodioxins/dibenzofurans and polybrominated diphenyl ethers from
open burning of domestic waste; environmental pollution due to e-waste & \cite{Mexico_1,Mexico_2,Mexico_3,Mexico_4,Mexico_5,Mexico_6,Mexico_7,Mexico_8}\tabularnewline
\hline 
MA & Only 37\% of the collected SW is disposed off in controlled landfills;
non-civic behaviors and deterioration of the environment have been
reported; BTEX, PAH and PCB are reported in different environmental
reservoirs. & \cite{Morocco_1,Morocco_2,Morocco_3,Morocco_4,Morocco_5}\tabularnewline
\hline 
MZ & \raggedright{}Heavy metals and organic chemicals, including pharmaceuticals,
in soils and waters. & \cite{Mozambique_1,Mozambique_2,Mozambique_3}\tabularnewline
\hline 
NG & \raggedright{}Heavy metals contamination and ecological risks from
municipal central dumpsite; contamination by heavy metals, VOC and
POP due to informal e-waste recycling; PAHs and PCB in groundwater
around waste dumpsites in South\nobreakdash-West Nigeria. & \cite{Nigeria_1,Nigeria_2,Nigeria_3,Nigeria_4,Nigeria_5,Nigeria_6,Nigeria_7}\tabularnewline
\hline 
PK & \raggedright{}Heavy metals pollution from diverse waste sources, including
e-waste, recognized as an emerging problem and medical waste incineration;
high levels of VOC, including BTEX in urban atmosphere; public health
problems from hospital solid waste mismanagement. & \cite{Pakistan_1,Pakistan_2,Pakistan_3,Pakistan_4,Pakistan_5,Pakistan_6,Pakistan_7,Pakistan_8}\tabularnewline
\hline 
PG & \raggedright{}Heavy metal water pollution in Depapre waters. & \cite{Papua_1}\tabularnewline
\hline 
SN & Dramatic problems with household waste collection; reported waste
contamination at different reservoirs by HM, mainly in the coast;
environmental issues with biomedical waste disposal; 18 children died
from a rapidly progressive central nervous system disease of unexplained
origin in a community involved in the recycling of used lead-acid
batteries; high levels of contamination with Pb in homes and soil
in surrounding areas dedicated to car battery informal recycling,
several children showed severe neurologic features of toxicity; high
levels of VOC contamination at different environmental places. & \cite{Senegal_1,Senegal_2,Senegal_3,Senegal_4,Senegal_5,Senegal_6,Senegal_7,Senegal_8,Senegal_9}\tabularnewline
\hline 
SL & \raggedright{}Heavy metals and POP contamination due to e-waste recycling;
high levels of exposition to dioxins and furans; mismanagement of
solid waste depositions with environmental and human health risk. & \cite{SierraLeone_1,SierraLeone_2,SierraLeone_3}\tabularnewline
\hline 
UZ & \raggedright{}Pollution by heavy metals with impact on human health;
contamination of soils with POP, particularly by PAHs. & \cite{Uzbekistan_1,Uzbekistan_2,Uzbekistan_3}\tabularnewline
\hline 
\end{longtable}

\section*{Countries/territories without EPI}

\begin{table}[H]
\begin{centering}
\begin{tabular}{|c|c|c|c|c|c|c|c|}
\hline 
\multicolumn{2}{|c|}{Type I} &  & \multicolumn{2}{c|}{Type II} &  & \multicolumn{2}{c|}{Type III}\tabularnewline
\hline 
country & $R_{i}$ &  & country & $R_{i}$ &  & country & $R_{i}$\tabularnewline
\hline 
\hline 
San Marino & 0.646 &  & Andorra & 0.461 &  & Monaco & 0.452\tabularnewline
\hline 
Liechtenstein & 0.616 &  & Puerto Rico & 0.461 &  & San Marino & 0.422\tabularnewline
\hline 
Monaco & 0.616 &  & Faroe Islands & 0.423 &  & Andorra & 0.368\tabularnewline
\hline 
Andorra & 0.596 &  & San Marino & 0.397 &  & Faroe Islands & 0.226\tabularnewline
\hline 
Guernsey & 0.586 &  & Monaco & 0.385 &  &  & \tabularnewline
\hline 
Isle of Man & 0.586 &  & Hong Kong & 0.372 &  &  & \tabularnewline
\hline 
Hong Kong & 0.566 &  & Liechtenstein & 0.372 &  &  & \tabularnewline
\hline 
Faroe Islands & 0.555 &  & New Caledonia & 0.359 &  &  & \tabularnewline
\hline 
New Caledonia & 0.545 &  & Guernsey & 0.446 &  &  & \tabularnewline
\hline 
Falkland Islands & 0.515 &  & Isle of Man & 0.333 &  &  & \tabularnewline
\hline 
St. Barthelemy & 0.495 &  & Jersey & 0.333 &  &  & \tabularnewline
\hline 
Jersey & 0.485 &  & Guernsey & 0.308 &  &  & \tabularnewline
\hline 
Palestine & 0.384 &  & Niue & 0.308 &  &  & \tabularnewline
\hline 
St. Kitts \& Nevis & 0.364 &  & Cook Islands & 0.256 &  &  & \tabularnewline
\hline 
Cook Islands & 0.333 &  & Falkland Islands & 0.243 &  &  & \tabularnewline
\hline 
Niue & 0.323 &  & Palestine & 0.159 &  &  & \tabularnewline
\hline 
Tuvalu & 0.313 &  &  &  &  &  & \tabularnewline
\hline 
\end{tabular}
\par\end{centering}
\caption{Values of the risk of congestion $R_{i}$ of waste types I-III for
countries/territories for which EPI are not reported. Therefore, we
cannot built the PEIWC for these countries/territories although they
can be at relatively HRIHDW.}

\end{table}

\section*{Waste categories in types IV-VII}

Waste of types IV-VII represents less than 0.001\% of the total volume
of waste traded in the world in the period 2003-2009. However, in
volume it still represents 2866.79 tonnes of waste traded across the
world: 821.92 tonnes (type IV), 295.90 tonnes (type V), 756.15 tonnes
(type VI) and 992.83 tonnes (type VII). The different categories of
the Basel Convention included in these four types are described in
Tables \ref{Type_IV}, \ref{Type_V}, \ref{Type_VI}, and \ref{Type_VII},
respectively.

\begin{table}[H]
\begin{centering}
\begin{tabular}{|c|>{\raggedright}p{10cm}|}
\hline 
category & description\tabularnewline
\hline 
\hline 
A1010 & Metal wastes and waste consisting of alloys of any of the following:
Antimony, Arsenic, Beryllium, Cadmium, Lead, Mercury, Selenium, Tellurium,
Thallium.\tabularnewline
\hline 
A1020 & Waste having as constituents or contaminants, excluding metal waste
in massive form, any of the following: Antimony; antimony compounds,
Beryllium; beryllium compounds, Cadmium; cadmium compounds, Lead;
lead compounds, Selenium; selenium compounds, Tellurium; tellurium
compounds\tabularnewline
\hline 
A1030 & Wastes having as constituents or contaminants any of the following:
Arsenic; arsenic compounds, Mercury; mercury compounds, Thallium;
thallium compounds\tabularnewline
\hline 
A1040 & Wastes having as constituents any of the following: Metal carbonyls,
hexavalent chromium compounds\tabularnewline
\hline 
A1050 & Galvanic sludges\tabularnewline
\hline 
A1060 & Waste liquors from the pickling of metals\tabularnewline
\hline 
A1070 & Leaching residues from zinc processing, dust and sludges, such as
jarosite, hematite, etc.\tabularnewline
\hline 
A1080 & Waste zinc residues, containing lead and cadmium in concentrations
sufficient to exhibit Annex III characteristics\tabularnewline
\hline 
A1090 & Ashes from the incineration of insulated copper wire\tabularnewline
\hline 
A1100 & Dusts and residues from gas cleaning systems of copper smelters\tabularnewline
\hline 
A1110 & Spent electrolytic solutions from copper electrorefining and electrowinning
operations\tabularnewline
\hline 
A1120 & Waste sludges, excluding anode slimes, from electrolyte purification
systems in copper electrorefining and electrowinning operations\tabularnewline
\hline 
A1130 & Spent etching solutions containing dissolved copper\tabularnewline
\hline 
A1140 & Waste cupric chloride and copper cyanide catalysts\tabularnewline
\hline 
A1150 & Precious metal ash from incineration of printed circuit boards\tabularnewline
\hline 
A1160 & Waste lead-acid batteries, whole or crushed\tabularnewline
\hline 
A1170 & Unsorted waste batteries. Waste batteries containing Annex I constituents
to an extent to render them hazardous\tabularnewline
\hline 
A1180 & Waste electrical and electronic assemblies or scrap containing components
such as accumulators and other batteries included on list A, mercury-switches,
glass from cathode-ray tubes and other activated glass and PCB capacitors,
or contaminated with Annex I constituents (e.g., cadmium, mercury,
lead, polychlorinated biphenyl) to an extent that they possess any
of the characteristics contained in Annex III\tabularnewline
\hline 
A1190 & Waste metal cables coated or insulated with plastics containing or
contaminated with coal tar, PCB, lead, cadmium, other organohalogen
compounds or other Annex I constituents to an extent that they exhibit
Annex III characteristics.\tabularnewline
\hline 
\end{tabular}
\par\end{centering}
\caption{Waste categories included in the Basel Convention which are grouped
in the type IV of wastes.}

\label{Type_IV}
\end{table}

\begin{table}[H]
\begin{centering}
\begin{tabular}{|c|>{\raggedright}p{10cm}|}
\hline 
category & description\tabularnewline
\hline 
\hline 
A2010 & Glass waste from cathode-ray tubes and other activated glasses\tabularnewline
\hline 
A2020 & Waste inorganic fluorine compounds in the form of liquids or sludges\tabularnewline
\hline 
A2030 & Waste catalysts\tabularnewline
\hline 
A2040 & Waste gypsum arising from chemical industry processes, when containing
Annex I constituents to the extent that it exhibits an Annex III hazardous
characteristic\tabularnewline
\hline 
A2050 & Waste asbestos (dusts and fibres)\tabularnewline
\hline 
A2060 & Coal-fired power plant fly-ash containing Annex I substances in concentrations
sufficient to exhibit Annex III characteristics\tabularnewline
\hline 
\end{tabular}
\par\end{centering}
\caption{Waste categories included in the Basel Convention which are grouped
in the type V of wastes.}

\label{Type_V}
\end{table}

\begin{table}[H]
\begin{centering}
\begin{tabular}{|c|>{\raggedright}p{10cm}|}
\hline 
category & description\tabularnewline
\hline 
\hline 
A3010 & Waste from the production or processing of petroleum coke and bitumen\tabularnewline
\hline 
A3020 & Waste mineral oils unfit for their originally intended use\tabularnewline
\hline 
A3030 & Wastes that contain, consist of or are contaminated with leaded anti-knock
compound sludges\tabularnewline
\hline 
A3040 & Waste thermal (heat transfer) fluids\tabularnewline
\hline 
A3050 & Wastes from production, formulation and use of resins, latex, plasticizers,
glues/adhesives\tabularnewline
\hline 
A3060 & Waste nitrocellulose\tabularnewline
\hline 
A3070 & Waste phenols, phenol compounds including chlorophenol in the form
of liquids or sludges\tabularnewline
\hline 
A3080 & Waste ethers\tabularnewline
\hline 
A3090 & Waste leather dust, ash, sludges and flours when containing hexavalent
chromium compounds or biocides\tabularnewline
\hline 
A3100 & Waste paring and other waste of leather or of composition leather
not suitable for the manufacture of leather articles containing hexavalent
chromium compounds or biocides\tabularnewline
\hline 
A3110 & Fellmongery wastes containing hexavalent chromium compounds or biocides
or infectious substances\tabularnewline
\hline 
A3120 & Fluff-light fraction from shredding\tabularnewline
\hline 
A3130 & Waste organic phosphorous compounds\tabularnewline
\hline 
A3140 & Waste non-halogenated organic solvents\tabularnewline
\hline 
A3150 & Waste halogenated organic solvents\tabularnewline
\hline 
A3160 & Waste halogenated or unhalogenated non-aqueous distillation residues
arising from organic solvent recovery operations\tabularnewline
\hline 
A3170 & Wastes arising from the production of aliphatic halogenated hydrocarbons
(such as chloromethane, dichloro-ethane, vinyl chloride, vinylidene
chloride, allyl chloride and epichlorhydrin)\tabularnewline
\hline 
A3180 & Wastes, substances and articles containing, consisting of or contaminated
with polychlorinated biphenyl (PCB), polychlorinated

terphenyl (PCT), polychlorinated naphthalene (PCN) or Polybrominated
biphenyl (PBB), or any other polybrominated analogues of these compounds,
at a concentration level of 50 mg/kg or more\tabularnewline
\hline 
A3190 & Waste tarry residues (excluding asphalt cements) arising from refining,
distillation and any pyrolitic treatment of organic materials\tabularnewline
\hline 
A3200 & Bituminous material (asphalt waste) from road construction and maintenance,
containing tar\tabularnewline
\hline 
\end{tabular}
\par\end{centering}
\caption{Waste categories included in the Basel Convention which are grouped
in the type VI of wastes.}

\label{Type_VI}
\end{table}

\begin{table}[H]
\begin{centering}
\begin{tabular}{|c|>{\raggedright}p{10cm}|}
\hline 
category & description\tabularnewline
\hline 
\hline 
A4010 & Wastes from the production, preparation and use of pharmaceutical
products\tabularnewline
\hline 
A4020 & Clinical and related wastes; that is wastes arising from medical,
nursing, dental, veterinary, or similar practices, and wastes generated
in hospitals or other facilities during the investigation or treatment
of patients, or research projects\tabularnewline
\hline 
A4030 & Wastes from the production, formulation and use of biocides and phytopharmaceuticals,
including waste pesticides and herbicides which are off-specification,
outdated, or unfit for their originally intended use\tabularnewline
\hline 
A4040 & Wastes from the manufacture, formulation and use of wood preserving

chemicals\tabularnewline
\hline 
A4050 & Wastes that contain, consist of or are contaminated with any of the
following: Inorganic cyanides, excepting precious-metal-bearing, residues
in solid form containing traces of inorganic cyanides, organic cyanides\tabularnewline
\hline 
A4060 & Waste oils/water, hydrocarbons/water mixtures, emulsions\tabularnewline
\hline 
A4070 & Wastes from the production, formulation and use of inks, dyes, pigments,
paints, lacquers, varnish\tabularnewline
\hline 
A4080 & Wastes of an explosive nature\tabularnewline
\hline 
A4090 & Waste acidic or basic solutions\tabularnewline
\hline 
A4100 & Wastes from industrial pollution control devices for cleaning of industrial
off-gases\tabularnewline
\hline 
A4110 & Wastes that contain, consist of or are contaminated with any of the
following: Any congenor of polychlorinated dibenzo-furan; Any congenor
of polychlorinated dibenzo-p-dioxin\tabularnewline
\hline 
A4120 & Wastes that contain, consist of or are contaminated with peroxides\tabularnewline
\hline 
A4130 & Waste packages and containers containing Annex I substances in concentrations
sufficient to exhibit Annex III hazard characteristics\tabularnewline
\hline 
A4140 & Waste consisting of or containing off specification or outdated chemicals
corresponding to Annex I categories and exhibiting Annex III hazard
characteristics\tabularnewline
\hline 
A4150 & Waste chemical substances arising from research and development or
teaching activities which are not identified and/or are new and whose
effects on human health and/or the environment are not known\tabularnewline
\hline 
A4160 & Spent activated carbon\tabularnewline
\hline 
\end{tabular}
\par\end{centering}
\caption{Waste categories included in the Basel Convention which are grouped
in the type VII of wastes.}

\label{Type_VII}
\end{table}

\section*{PEIWS analysis of wastes types IV-VII}

\begin{figure}[H]
\begin{centering}
\includegraphics[width=12cm]{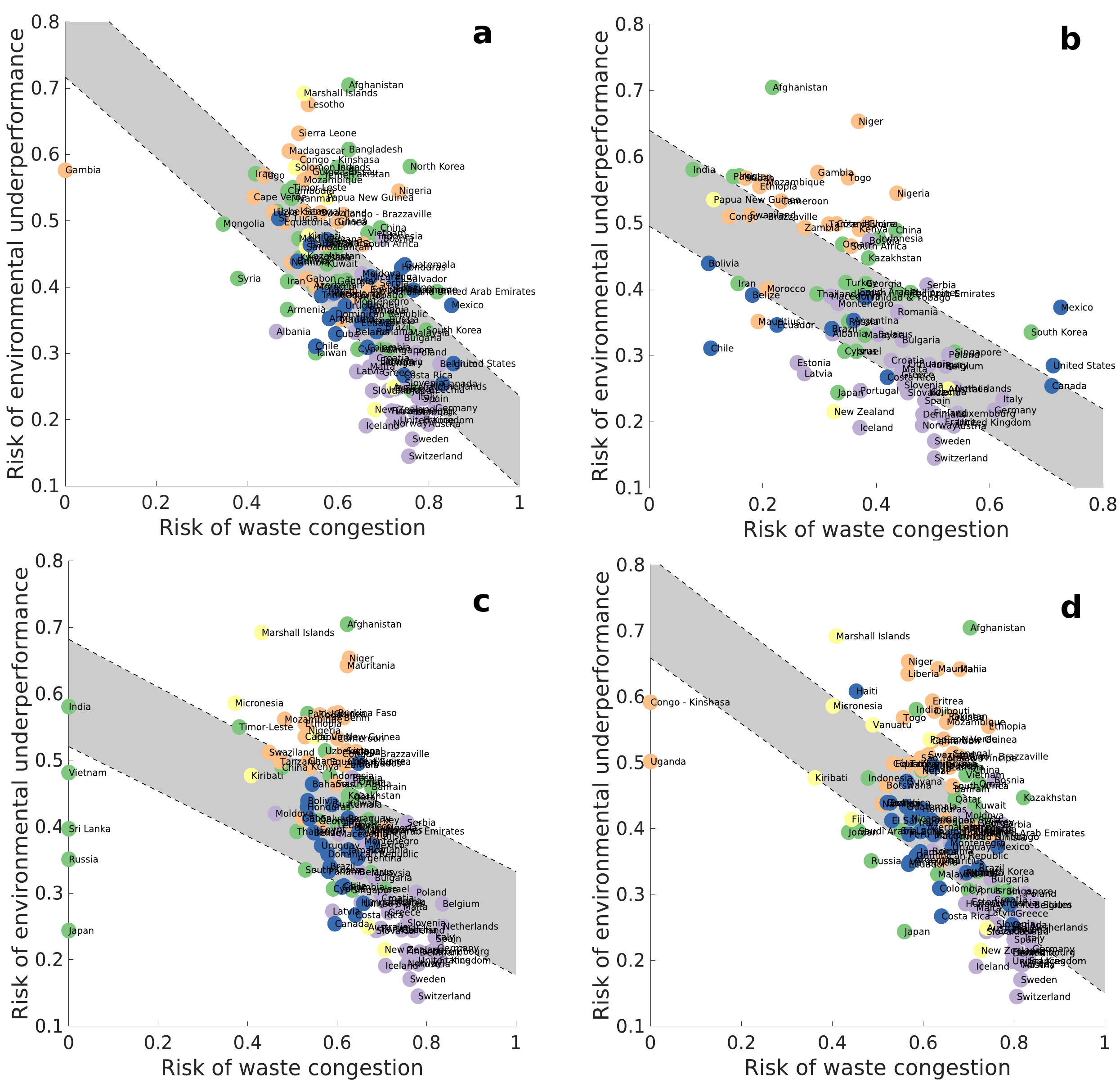}
\par\end{centering}
\caption{\textbf{PEIWC} \textbf{of wastes types IV-VII}. (a)-(d) Illustration
of the PEIWC for wastes of types IV-VII, respectively. The risks of
waste congestion are calculated from the simulated dynamics using
a fractional logistic model described in Methods. The
index of risk of environmental underperformance is obtained from the
Yale University environmental performance index (EPI). Nodes are colored by the continent to which thecountry belongs to:  blue (Americas), purple (Europe), yellow (Africa), green (Asia)}
\label{PEIWS_IV_VII}
\end{figure}

Following the same procedure described in Methods we build the PEIWS
of the four types of waste IV-VII, which are illustrated in Fig. \ref{PEIWS_IV_VII}.
Using the same approach as for the waste types I-III we identify these
countries at HRIHDW. In total in the four types of waste there are
29 countries at HRIHDW, 22 of which coincide with countries previously
identified at HRIHDW for waste types I-III. The new countries at HRIHDW,
i.e., not identified for types I-III, are Kazakhstan, Mongolia, C\^ote
d'Ivoire, Saudi Arabia, Tanzania, Kenya and Oman. Wastes of types
IV and VI are the ones with the largest number of countries at HRIHDW
with 15 and 12, respectively, while types V and VII have 8 and 9 countries
at HRIHDW, respectively. By continents, Africa is again the one having
more countries at HRIHDW with 12, followed by Asia (9) and then Middle
East and Europe with 4 each.

\pagebreak